\documentclass[aps,twocolumn,showpacs,pra,twoside,amssymb,amsmath,latexsym]{revtex4}
\usepackage{graphicx}
\newtheorem{lem}     {Lemma}

\def\choose#1#2{\genfrac{(}{)}{0pt}{}{#1}{#2}}
\def\Tr{\mathop{\mathbf{Tr}}\nolimits}
\def\UMP{\mathop{\rm UMP}\nolimits}
\def\LR{\mathop{\rm LR}\nolimits}
\def\new{\mathop{\rm mod}\nolimits}

\def\ket#1{|#1\rangle}

\def\cA{{\cal A}}
\def\cB{{\cal B}}
\def\cH{{\cal H}}
\def\fid{F}
\def\complex{\mathbb{C}}
\def\real{\mathbb{R}}
\begin{document}
\title{Statistical analysis on testing 
of an entangled state\\
based on Poisson distribution framework}
\author{Masahito Hayashi$^{\ast,\dag}$,
Akihisa Tomita$^{\ast,\ddag}$, and Keiji Matsumoto$^{\S,\ast}$}
\affiliation{$^{\ast}$ERATO-SORST Quantum Computation and Information Project, 
Japan Science and Technology Agency (JST), Tokyo 113-0033, Japan \\
$^{\dag}$Superrobust Computation Project,
Information Science and Technology Strategic Core (21st Century COE by 
MEXT)\\
Graduate School of Information Science and Technology,
The University of Tokyo\\
7-3-1, Hongo, Bunkyo-ku, Tokyo, 113-0033, Japan\\
$^{\ddagger}$Fundamental Research Laboratories, NEC, Tsukuba 305-8501, Japan\\
$^{\S}$National Institute of Informatics, Hitotsubashi, Chiyoda-ku, Tokyo 101-8430, Japan}
\begin{abstract}
A hypothesis testing scheme for entanglement has been formulated
based on the Poisson distribution framework
instead of the POVM framework. 
Three designs were proposed to test the entangled states in this framework.
The designs were evaluated in terms of the asymptotic variance.
It has been shown that the optimal time allocation between the coincidence
and anti-coincidence measurement
bases improves the conventional testing method. 
The test can be further improved by
optimizing the time allocation between the anti-coincidence bases.
\end{abstract}
\maketitle

\section{Introduction}\label{s1}
Entangled states are an essential resource for various 
quantum information processings\cite{Bennett93,Briegel98}. 
Hence, it is required to generate
maximally entangled states.
However, for a practical use,
it is more essential to guarantee the quality of generated
entangled states.
Statistical hypothesis testing is a standard method
for guaranteeing the quality of industrial products.
Therefore, it is 
much needed to establish the method 
for statistical testing of maximally entangled states.

Quantum state estimation and 
quantum state tomography 
are known as the method of identifying the unknown 
state\cite{Selected,helstrom,holevo}.
Quantum state tomography \cite{WJEK99} has been recently applied to obtain full
information of the $4 \times 4$ density matrix. 
However, if the purpose is testing of entanglement, 
it is more economical to concentrate on checking the degree of 
entanglement.
Such a study has been done by 
Tsuda et al \cite{TMH05} 
as optimization problems of POVM.
However, an implemented quantum measurement cannot be regarded as
an application of a POVM to a single particle system or
a multiple application of a POVM to single particle systems.
In particular, in quantum optics,
the following measurement is often realized, which is not described by 
a POVM on a single particle system.
The number of generated particles is probabilistic.
We prepare a filter corresponding to a projection $P$,
and detect the number of particle passing through the filter.
If the number of generated particles 
obeys a Poisson distribution,
as is mentioned in Section \ref{s2},
the number of detected particles obeys another 
Poisson distribution whose average is given by the 
density and the projection $P$.

In this kind of measurements,
if any particle is not detected,
we cannot decide whether a particle is not generated or
it is generated but does not pass through the filter.
If we can detect the number of generated particles as well as
the number of passing particles,
the measurement can be regarded as
the multiple application of 
the POVM $\{P,I-P\}$.
In this case, the number of 
applications of the POVM is the variable corresponding to 
the number of generated particles.
Also, we only can detect the empirical distribution.
Hence, our obtained information almost 
discuss by use of the POVM $\{P,I-P\}$.

However, if it is impossible to distinguish the two events
by some imperfections,
it is impossible to reduce 
the analysis of our obtained information to the analysis of POVMs.
Hence, it is needed to analyze 
the performance of the estimation and/or the hypothesis testing
based on the Poisson distribution describing the 
number of detected particles.
If we discuss the ultimate bound 
of the accuracy of the estimation and/or the hypothesis testing,
we do not have to treat such imperfect measurements.
Since several realistic measurements have 
such imperfections,
it is very important to optimize our measurement among 
such a class of imperfect measurements.

In this paper, our measurement is restricted to 
the detection of the number of the particle passing through the 
filter corresponding to a projection $P$.
We apply this formulation to the testing of maximally entangled states
on two qubit systems (two-level systems),
each of which is spanned by
two vectors $\vert H\rangle$ and $\vert V\rangle$.
Since the target system is a bipartite system,
it is natural to restrict to our measurement to 
local operations and classical communications (LOCC).
In this paper, for a simple realization,
we restrict our measurements
to the number of the simultaneous detections at the 
both parties of the particles passing through 
the respective filters.
We also restrict the total measurement time $t$,
and optimize the allocation of the time for each 
filters at the both parties.

As our results,
we obtain the following characterizations.
If the average number of the generated particles is known,
our choice is counting the coincidence events
or the anti-coincidence events.
When the true state is close to the target 
maximally entangled state 
$\vert \Phi^{(+)} \rangle:=\frac{1}{\sqrt{2}}(\vert HH\rangle+\vert VV\rangle)$
(that is, the fidelity between these is greater than $1/4$),
the detection of anti-coincidence events 
is better than that of coincidence events.
This result implies that
the indistinguishability 
between the coincidence events and the non-generation event 
loses less information
than that 
between the anti-coincidence events and the non-generation event.

This fact also holds even if we treat this problem
taking into account the effect of dark counts.
In this discussion, 
in order to remove the bias concerning 
the direction of the difference,
we assume the equal time allocation 
among the vectors 
$\{\vert HV\rangle, \vert VH\rangle, \vert DX\rangle, \vert XD\rangle, 
\vert RR\rangle, \vert LL\rangle \}$,
which corresponds to the anti-coincidence events,
and that among the vectors 
$\{\vert HH\rangle, \vert VV\rangle, \vert DD\rangle, \vert XX\rangle, 
\vert RL\rangle, \vert LR\rangle \}$,
which corresponds to the coincidence events,
where
$\vert D\rangle:=
\frac{1}{2}(\vert H\rangle+ \vert V\rangle)$,
$\vert X\rangle:=
\frac{1}{2}(\vert H\rangle- \vert V\rangle)$,
$\vert R\rangle:=
\frac{1}{2}(\vert H\rangle+i\vert V\rangle)$,
$\vert L\rangle:=
\frac{1}{2}(\vert H\rangle-i\vert V\rangle)$.
Indeed, Barbieri et al \cite{BMNMDM03} 
proposed to detect 
the anti-coincidence events for 
measuring an entanglement witness,
they did not prove the superiority of detecting
the anti-coincidence events 
in the framework of mathematical statistics.

However, the average number of 
the generated particles is usually unknown.
In this case, we cannot estimate how close the true state 
is to the target maximally entangled state
from the detection of anti-coincidence events.
Hence, we need to count the coincidence events
as additional information.
in order to resolve this problem,
we usually use the equal allocation between 
anti-coincidence events and coincidence events
in the visibility method, which is a conventional method
for checking the entanglement.
However, since we measure the coincidence events and the anti-coincidence 
events based on one or two bases in this method, 
there is a bias concerning the direction of the difference.
In order to remove this bias,
we consider the detecting method with the equal time allocation 
among all vectors
$\{\vert HV\rangle, \vert VH\rangle, \vert DX\rangle, \vert XD\rangle, 
\vert RR\rangle, \vert LL\rangle \}$
and 
$\{\vert HH\rangle, \vert VV\rangle, \vert DD\rangle, \vert XX\rangle, 
\vert RL\rangle, \vert LR\rangle \}$,
and call it the modified visibility method.

In this paper, we also examine the detection of the total flux,
which can be realized by detecting the particle without the filter.
We optimize the time allocation among 
these three detections.
We found that 
the optimal time allocation depends on 
the fidelity between the true state and the target maximally entangled state.
If our purpose is estimating the fidelity $\fid$,
we cannot directly apply the optimal time allocation.
However, the purpose is testing whether the fidelity $\fid$ is greater than 
the given threshold $\fid_0$,
the optimal allocation at $\fid_0$ gives 
the optimal testing method.

If the fidelity $\fid$ is less than a critical value,
the optimal allocation is given by 
the allocation between 
the anti-coincidence vectors and 
the coincidence vectors (the ratio depends on $\fid$.)
Otherwise, it is given by the allocation only between 
the anti-coincidence vectors and the total flux.
This fact is valid even if the dark count exists.
If the dark count is greater than a certain value,
the optimal time allocation is always given by 
the allocation between 
the anti-coincidence vectors and 
the coincidence vectors.

Further, we consider the optimal allocation 
among anti-coincidence vectors 
when the average number of generated particles.
The optimal allocation depends on the direction of the difference
between the true state and the target state.
Since the direction is usually unknown,
this optimal allocation dose not seems useful.
However, by adaptively deciding the optimal time allocation,
we can apply the optimal time allocation.
We propose to apply this optimal allocation 
by use of the two-stage method.
Further, taking into account the complexity of testing methods
and the dark counts,
we give a testing procedure of entanglement 
based on the two-stage method.
In addition, proposed designs of experiments were demonstrated 
by Hayashi et al. \cite{HSTMTJ} in two photon pairs generated by 
spontaneous parametric down conversion (SPDC).

In this article, we reformulate the hypothesis testing to be applicable to
the Poisson distribution framework,
and demonstrate the effectiveness 
of the optimized time allocation in the entanglement test.  
The construction of this article is following. 
Section \ref{s2} defines 
the Poisson distribution framework and gives
the hypothesis scheme for the entanglement.
Section \ref{s3} gives the mathematical formulation concerning
statistical hypothesis testing.
Sections \ref{s4} and \ref{s5} give 
the fundamental properties of the hypothesis testing:
section \ref{s4} introduces the likelihood ratio test and its modification, 
and
section \ref{s5} gives the asymptotic theory of the hypothesis testing.
Sections \ref{s6}-\ref{s9} are devoted to the designs of the time allocation 
between the coincidence and anti-coincidence bases: 
section \ref{s6} defines the modified visibility method,
 section \ref{s7} optimize the time allocation, 
when the total photon flux $\lambda$ is unknown, 
section \ref{s8} gives the results with known $\lambda$, 
and section \ref{s9} compares the designs in terms of the asymptotic variance.
Section \ref{s10} gives further improvement by optimizing the time allocation
between the anti-coincidence bases.
Appendices give the detail of the proofs used in the optimization.

\section{Hypothesis Testing scheme for entanglement
in Poisson distribution framework}\label{s2}
Let $\cH$ be the Hilbert space of our interest, and 
$P$ be the projection corresponding to our filter.
If we assume generation process on each time 
to be identical but individual,
the total number $n$ of generated particles 
during the time $t$ obeys the Poisson distribution
${\rm Poi}(\lambda t)(n):= e^{-\lambda t}\frac{(\lambda t)^n}{n !}$.
Hence, when the density of the true state is $\sigma$,
the probability of the number $k$ of detected particles 
is given as
\begin{align}
&\sum_{n=0}^\infty 
{\rm Poi}(\lambda t)(n)
\choose{n}{k}
(\Tr P \sigma)^k
(1- \Tr P \sigma)^{n-k}\nonumber \\
=&
{\rm Poi}(\lambda t
\Tr P \sigma )(k).\label{5-8-1}
\end{align}

\begin{figure}[htbp]
     \begin{center}
  \includegraphics*[width=8cm]{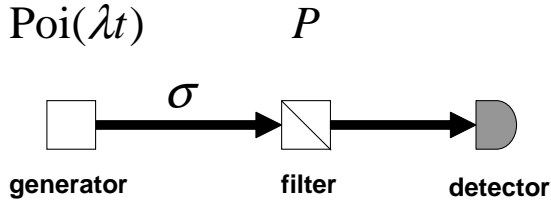}
  \end{center}
\caption{Experimental scheme in Poisson distribution framework}
   \label{scheme}
\end{figure}
In fact, 
if we treat the Fock space generated by $\cH$ 
instead of the single particle system $\cH$,
this measurement can be described by a POVM.
However, since this POVM dooes not have a simple form,
it is suitable to treat this measurement in the form (\ref{5-8-1}).

Further, if we errorly detect the $k'$ particles 
with the probability ${\rm Poi}(\delta t)(k')$,
the probability of the number $k$ of detected particles 
is equal to 
\begin{align*}
&\sum_{k'=0}^k
{\rm Poi}(\lambda t
\Tr P \sigma )(k-k')
+
{\rm Poi}(\delta t)(k')\\
=& 
{\rm Poi}
((\lambda \Tr P \sigma +\delta )t)(k).
\end{align*}
This kind of incorrect detection 
is called dark count.
Further, since we consider the bipartite case, i.e., the case where 
$\cH= \complex^2 \otimes \complex^2$,
we assume that our projection $P$ has the separable form
$P_1 \otimes P_2$.

In this paper, 
under the above assumption, we discuss the hypothesis testing 
when 
the target state is the maximally entangled $\vert \Phi^{(+)} \rangle $ state
while Usami et al.\cite{Usami} 
discussed the state estimation under this assumption.
Here we measure the degree of entanglement by the fidelity 
between the generated state and the target state:
\begin{equation}
\fid = \langle \Phi^{(+)} \vert \sigma \vert \Phi^{(+)} \rangle. 
\label{Fidelity}
\end{equation}
The purpose of the test is to guarantee that the state is sufficiently 
close to the maximally entangled state with a certain significance.
That is, we are required to disprove 
that the fidelity $\fid$ is less than a threshold $\fid_0$ 
with a small error probability.
In mathematical statistics, this situation is 
formulated as hypothesis testing;
 we introduce the null hypothesis $H_0$ that
entanglement is not enough and 
the alternative $H_1$ that the entanglement is enough:
\begin{align}
H_0:\fid \le \fid_0 \hbox{ v.s. }
H_1:\fid > \fid_0,\label{5-5-2}
\end{align}
with a threshold $\fid_0$.

Visibility is an indicator of 
entanglement commonly used in the experiments,
and is calculated as follows:
first, A's measurement vector $\ket{x_A}$ is fixed,
then the measurement $\ket{x_A,y_B}$ is performed by rotating 
B's measurement vector $\ket{x_B}$ to obtain the maximum and minimum 
number of the counts, $n_{max}$ and $n_{min}$.
We need to make the measurement with at least two bases of A in order to
exclude the possibility of the classical correlation. 
We may choose the two bases 
$\{|H\rangle,|V\rangle\}$ and 
$\{|D\rangle,|X\rangle\}$ as $\ket{x_A}$, 
for example.
Finally, the visibility is given 
by the ratio between $n_{max}-n_{min}$ and $n_{max}+n_{min}$
with the respective A's measurement basis $\ket{x_A}$.
However, our decision will contain a bias, 
if we choose only two bases as A's measurement basis $\ket{x_A}$.
Hence, we cannot estimate the fidelity between 
the target maximally entangled state and the given state
in a statistically proper way from the visibility. 

Since the equation
\begin{align}
&|HH\rangle\langle HH|+|VV\rangle\langle VV|+
|DD\rangle\langle DD|\nonumber \\
&+|XX\rangle\langle XX|+
|RL\rangle\langle RL|+|LR\rangle\langle LR|\nonumber \\
=&
2|\Phi^{(+)}\rangle\langle\Phi^{(+)}|+ I\label{2-12}
\end{align}
holds,
we can estimate the fidelity 
by measuring the sum of
the counts of 
the following vectors:
$|HH\rangle, |VV\rangle, |DD\rangle, |XX\rangle, 
|RL\rangle,$ and $|LR\rangle$,
when $\lambda$ is known\cite{BMNMDM03,TMH05}.
This is because
the sum $n_1:= n_{HH}+n_{VV}+ n_{DD}+ n_{XX}+n_{RL}+n_{LR}$ 
obeys the Poisson distribution with the expectation value 
$(\lambda \frac{1+ 2\fid }{6}+\delta)t_1$,
where the measurement time for each vector is $\frac{t_1}{6}$. 
We call these vectors the coincidence vectors because
these correspond to the coincidence events.

However, since the parameter $\lambda$ is usually unknown,
we need to perform another measurement on different vectors 
to obtain
additional information. 
Since 
\begin{align}
&|HV\rangle\langle HV|+|VH\rangle\langle VH|+|XD\rangle\langle XD|\nonumber \\
&+|DX\rangle\langle DX|+|RR\rangle\langle RR|+|LL\rangle\langle LL|\nonumber \\
=&
2I - 2|\Phi^{(+)}\rangle\langle\Phi^{(+)}|
\end{align}
also holds, 
we can estimate the fidelity 
by measuring the sum of
the counts of 
the following vectors:
$|HV\rangle, |VH\rangle, |DX\rangle, |XD\rangle, 
|RR\rangle$, and $|LL\rangle$.
The sum $n_2:= n_{HV}+ n_{VH}+ n_{DX}+ n_{XD}+ n_{RR}+ n_{LL}$
obeys the Poisson distribution 
$\rm Poi((\lambda \frac{2- 2\fid }{6}+\delta)t_2)$,
where the measurement time for each vector is $\frac{t_2}{6}$. 
Combining the two measurements, we can estimate the fidelity 
without the knowledge of $\lambda$.
We call these vectors the anti-coincidence vectors because
these correspond to the anti-coincidence events.

We can also consider different type of measurement 
on $\lambda$.
If we prepare our device to detect all photons, i.e.,
the case where the projection is $I \otimes I$,
the detected number $n_3$ obeys the distribution
$\rm Poi((\lambda+\delta) t_3$) with the measurement time $t_3$. 
We will refer to it as the total flux measurement.
In the following, we consider the best time allocation for estimation
and test on the fidelity, 
by applying methods of mathematical statistics. 
We will assume that $\lambda$ is known or
estimated from the detected number $n_3$.

\section{Hypothesis testing for probability distributions}\label{s3}
\subsection{Formulation}
In this section, we review the fundamental knowledge
of hypothesis testing for probability distributions\cite{lehmann}. 
Suppose that a random variable $X$ is 
distributed according to
a probability measure $P_\theta$
identified by the unknown parameter $\theta$.
We also assume that
the unknown parameter $\theta$ 
belongs to one of 
mutually disjoint sets $\Theta_0$ and $\Theta_1$.
When we want to guarantee that
the true parameter $\theta$ belongs to the set $\Theta_1$
with a certain significance,
we choose the null hypothesis $H_0$ and 
the alternative hypothesis $H_1$ as
\begin{equation}
	\label{eq:hypo}
	H_0:\theta\in\Theta_0
	\mbox{ versus }
	H_1:\theta\in\Theta_1.
\end{equation}
Then, our decision method is described by 
a test, which is described as a function
$\phi(x)$ taking values in $\{0,1\}$;
$H_0$ is rejected if $1$ is observed,
and
$H_0$ is not rejected if $0$ is observed.
That is, we make our decision only when $1$ is observed,
and do not otherwise.
This is because
the purpose is accepting $H_1$ by rejecting $H_0$
with guaranteeing the quality of our decision,
and is not rejecting $H_1$ nor accepting $H_1$.
Therefore, we call the region $\{x|\phi(1)=1\}$ the 
rejection region.
The test $\phi$ can be defined by the rejection region.
In fact, we choosed the hypothesis that the fidelity is less than the 
given threshold $\theta_0$ as the null hypothesis $H_0$ in Section \ref{s2}.
This formulation is natural because 
our purpose is guaranteeing that the fidelity is not less 
than the given threshold $\theta_0$.

From theoretical viewpoint, we often consider randomized tests,
in which we probabilistically make the decision for a given data.
Such a test is given by a function $\phi$ mapping to the interval $[0,1]$.
When we observe the data $x$,
$H_0$ is rejected with the probability $\phi(x)$.
In the following, we treat randomized tests as well as 
deterministic tests.

In the statistical hypothesis testing,
we minimize error probabilities of the test $\phi$.
There are two types of errors.
The type one error is the case where
$H_0$ is rejected though it is true.
The type two error is the converse case,
$H_0$ is accepted though it is false.
Hence, the type one error probability is given 
$P_\theta(\phi)$ $(\theta \in \Theta_0)$, and
the type two error probability is given 
$1-P_{\theta'}(\phi)$ $(\theta' \in \Theta_1)$, where
\begin{eqnarray*}
	P_\theta(\phi)=
	\int\phi(x)d P_\theta(x).
\end{eqnarray*}
It is in general impossible to minimize both
$P_\theta(\phi)$ and $1-P_{\theta'}(\phi)$ simultaneously
because of a trade-off relation between them.
Since we make our decision 
with guaranteeing its quality 
only when $1$ is observed,
it is definitively required that
the type one error probability $P_\theta(\phi)$
is less than a certain constant $\alpha$.
For this reason,
we minimize the type two error probability $1-P_{\theta'}(\phi)$
under the condition $P_{\theta}(\phi)\le\alpha$.
The constant $\alpha$ in the condition is called 
the risk probability,
which guarantees the quality of our decision.
If the risk probability is large enough,
our decision has less reliability.
Under this constraint for the risk probability,
we maximize the probability to reject 
the hypothesis $H_0$ when the true parameter is $\theta'\in \Theta_1$.
This probability is given as $P_\theta(\phi)$,
and is called the power of $\phi$.
Hence, a test $\phi$ of the risk probability $\alpha$
is said to be most powerful (MP) at $\theta'\in\Theta_1$
if $P_{\theta'}(\phi)\ge P_{\theta'}(\psi)$ holds
for any test $\psi$ of the risk probability $\alpha$.
Then, a test is said to be Uniformly Most Powerful (UMP)
if it is MP at any $\theta'\in\Theta_1$.

\subsection{p-values} \label{pval}
In the hypothesis testing, 
we usually fixed our test before applying 
it to data.
However, we sometimes focus on the 
minimum risk probability 
among tests in a class $\tilde{T}$
rejecting  
the hypothesis $H_0$ with a given data.
This value is called the p-value, which depends on the observed data $x$
as well as the subset $\Theta_0$ to be rejected.

In fact, in order to define the p-value,
we have to fix a class $T$ of tests.
Then, 
for $x$ and $\Theta_0$, p-value is defined as
\begin{align}
\min_{\phi \in T: \phi(x)=1}
\max_{\theta \in \Theta_0}
P_{\theta} (\phi).
\end{align}
Since the p-value expresses the risk for rejecting the hypothesis $H_0$,
Hence, this concept is useful
for comparison among several designs of experiment.

Note that if we are allowed to choose any function $\phi$ as a test,
the above minimum is attained by the function $\delta_x$:
\begin{align}
\delta_x(y)= \left\{
\begin{array}{ll}
0 & \hbox{ if } y \neq x \\
1 & \hbox{ if } y = x .
\end{array}
\right.
\end{align}
In this case, the p-vale is 
$\max_{\theta \in \Theta_0}P_{\theta} (x)$.
However, the function $\delta_x$ is unnatural as a test.
Hence, we should fix a class of tests to define p-value.

\section{Likelihood Test}\label{s4}
\subsection{Definition}
In mathematical statistics, the likelihood ratio tests 
is often used as a class of standard tests\cite{lehmann}.
This kind of tests often provide
the UMP test in some typical cases.
When both $\Theta_0$ and $\Theta_1$ consist of single elements
as $\Theta_0=\{\theta_0\}$ and $\Theta_1=\{\theta_1\}$,
the likelihood ratio test
$\phi_{\LR,r}$ is defined as
\[
	\phi_{\LR,r}(x):=
	\begin{cases}
	0 & {\rm if }\
	P_{\theta_0}(x)/P_{\theta_1}(x)
	\ge r,
	\cr
	1 & {\rm if }\
	P_{\theta_0}(x)/P_{\theta_1}(x)
	< r
	\end{cases}
\]
where $r$ is a constant,
and the ratio
$P_{\theta_0}(x)/P_{\theta_1}(x)$ is called the likelihood ratio.
From the definition, any test $\phi$ satisfies
\begin{align}
(r P_{\theta_1} - P_{\theta_0})(\phi_{\LR,r})\ge 
(r P_{\theta_1} - P_{\theta_0})(\phi).
\end{align}

When a likelihood ratio test $\phi_{\LR,r}$ satisfies
\begin{align}
	\alpha=
	P_{\theta_0}(\phi_{\LR,r}),
\end{align}
the test $\phi_{\LR,r}$ is MP of level $\alpha$.
Indeed, when a test $\phi$ satisfies
$P_{\theta_0}(\phi) \le \alpha$,
\begin{align*}
&- \alpha + r P_{\theta_1}(\phi)
= - P_{\theta_0}(\phi)+ r P_{\theta_1}(\phi) \\
\le &- P_{\theta_0}(\phi_{\LR,r})+ r P_{\theta_1}(\phi_{\LR,r})
= - \alpha + r P_{\theta_1}(\phi_{\LR,r}).
\end{align*}
Hence, $1- P_{\theta_1}(\phi)\ge 1- P_{\theta_1}(\phi_{\LR,r})$.
This is known as Neyman-Pearson's fundamental lemma\footnote{}.

The likelihood ratio test is generalized to the
cases where $\Theta_0$ or $\Theta_1$
has at least two elements as
\[
	\phi_{\LR,r}(x):=
	\begin{cases}
	0 & {\rm if }\
	\frac
	{\sup_{\theta\in\Theta_0}P_\theta(x)}
	{\sup_{\theta\in\Theta_1}P_\theta(x)}
	\ge r,
	\cr
	1 & {\rm if }\
	\frac
	{\sup_{\theta\in\Theta_0}P_\theta(x)}
	{\sup_{\theta\in\Theta_1}P_\theta(x)}
	< r.
	\end{cases}
\]
Usually, in order to guarantee a small risk probability,
the likelihood ratio $r$ is choosed as $r <1$.

\subsection{Monotone Likelihood Ratio Test}
In cases where the hypothesis is one-sided,
that is,
the parameter space $\Theta$ is an interval of $\Bbb R$
and
the hypothesis is given as
\begin{equation}
	\label{eq:hypo-int}
	H_0:\theta\ge\,\theta_0
	\mbox{ versus }
	H_1:\theta < \,\theta_0
	,
\end{equation}
we often use so-called interval tests
for its optimality under some conditions
as well as for its naturalness.

When 
the likelihood ratio $P_\theta(x)/P_\eta(x)$ 
is monotone increasing concerning $x$ for any $\theta,\eta$
such that $\theta > \eta$,
the likelihood ratio is called monotone.
In this case, the likelihood ratio test $\phi_{\LR,r}$ 
between $P_{\theta_0}$ and $P_{\theta_1}$ 
is UMP of level $\alpha:= P_{\theta_0}(\phi_{\LR,r})$,
where $\theta_1$ is an arbitrary element
satisfying $\theta_1 < \theta_0$.

Indeed, many important examples satisfy this condition.
Hence, it is convenient to give its proof here.

From the monotonicity,
the likelihood ratio test $\phi_{\LR,r}$ 
has the form
\begin{align}
\phi_{\LR,r}(x)= 
\left\{
\begin{array}{ll}
1 & x  < x_0 \\
0 & x \ge x_0
\end{array}
\right.\label{2-14-1}
\end{align}
with a threshold value $x_0$.
Since the monotonicity implies
$P_{\theta_0}(\phi_{\LR,r})\ge P_{\theta}(\phi_{\LR,r})$ for any 
$\theta\in \Theta_0$,
it follows from Neyman Pearson Lemma that
the likelihood ratio test $\phi_{\LR,r}$ is MP of level $\alpha$.
From (\ref{2-14-1}),
the likelihood ratio test 
$\phi_{\LR,r}$ is also a likelihood ratio test between 
$P_{\theta_0}$ and $P_{\eta}$,
where $\eta$ is another element 
satisfying $\eta < \theta_0$.
Hence, the test $\phi_{\LR,r}$ is also MP of level $\alpha$.

From the above discussion,
it is suitable to treat p-value based on the class of likelihood ratio tests.
In this case, when we observe $x_0$,
the p-value is equal to
\begin{align}
\int_{-\infty}^{x_0} P_{\theta_0}(d x).
\end{align}


\subsection{One-Parameter Exponential Family}\label{2-7-5}
In mathematical statistics,
exponential families are known as
a class of typical statistical models\cite{A-N}.
A family of probability distributions
$\{P_\theta|\theta\subset \Theta\}$
is called an exponential family
when there exists a random variable $x$ such that
\begin{align}
	P_\theta(x):=P_0(x)
	\exp(\theta x+g(\theta))
	,\label{2-17-3}
\end{align}
where 
$g(\theta):= -\log \int \exp(\theta x)P_0(d x)$.

It is known that this class of families includes, 
for example,
the Poisson distributions, normal distributions, binomial distributions,
etc.
In this case,
the likelihood ratio 
$\frac{\exp(\theta_0 x+g(\theta_0))}{\exp(\theta_1 x+g(\theta_1))}
= \exp((\theta_0-\theta_1) x+g(\theta_0)-g(\theta_1))$
is monotone concerning $x$ for $\theta_0 > \theta_1$.
Hence, the likelihood ratio test is UMP
in the hypothesis (\ref{eq:hypo-int}).
Note that this argument is valid even if we choose a different parameter 
if the family has a parameter satisfying (\ref{2-17-3}).

For example, in the case of the normal distribution 
$P_\theta(x)=
\frac{1}{\sqrt{2\pi V}}e^{- \frac{(x-\theta)^2}{2V}}
=
\frac{1}{\sqrt{2\pi V}}
e^{- \frac{x^2}{2V}+ \frac{\theta x}{V} - \frac{\theta^2}{2V}}$,
the UMP test $\phi_{\UMP,\alpha}$ of the level $\alpha$
is given as
\begin{align}
\phi_{\UMP,\alpha}(x):=
\left\{
\begin{array}{ll}
1 & \hbox{ if } x <   \theta_0 - \epsilon_\alpha\sqrt{V} \\
0 & \hbox{ if } x \ge \theta_0 - \epsilon_\alpha\sqrt{V},
\end{array}
\right.
\end{align}
where
\begin{align*}
\Phi(-\epsilon_\alpha)=\alpha ,\quad
\Phi(\epsilon)= \int^{\epsilon}
_{-\infty} \frac{1}{\sqrt{2\pi}}
e^{-\frac{x^2}{2}}dx.
\end{align*}

The $n$-trial binomial distributions $
P_p^n(k)= \choose{n}{k}(1-p)^{n-k}p^k$
are also an exponential family
because 
another parameter 
$\theta:=\log \frac{p}{1-p}$ 
satisfies that
$P_p^n(k)= 
\choose{n}{k}
\frac{1}{2^n}
e^{\theta k + n \log \frac{e^\theta}{1+e^\theta}}$.
Hence, in the case of 
the $n$-trial binomial distribution,
the UMP test $\phi_{\UMP,\alpha}^n$ of the level $\alpha$
is given as the randomized likelihood ratio test:
\begin{align}
\phi_{\UMP,\alpha}(k):=
\left\{
\begin{array}{ll}
1 & \hbox{ if } x <   k_0 \\
\gamma & \hbox{ if } x = k_0\\
0 & \hbox{ if } x > k_0
\end{array}
\right. \label{5-4-1}
\end{align}
where
$k_0$ is the maximum value $k'$ satisfying
$\alpha \ge\sum_{k=0}^{k'-1}\choose{n}{k}(1-\theta)^{n-k}\theta^k$,
and $\gamma$ is defined as
\begin{align}
\alpha 
= \gamma \choose{n}{k_0}(1-\theta)^{n-k_0}\theta^{k_0}
+ \sum_{k=0}^{k_0-1}\choose{n}{k}(1-\theta)^{n-k}\theta^k.
\end{align}
Therefore, when $k$ is observed, the p-value is 
$\sum_{k'=0}^k\choose{n}{k'}(1-\theta_0)^{n-k'}\theta_0^{k'}$.

When $n$ is sufficiently large,
the distribution $P_\theta^n(k)$
can be approximated by the normal distribution
with variance $n(1-\theta)\theta$.
Hence, the UMP test $\phi_{\UMP,\alpha}^n$ of the level $\alpha$
is approximately given as
\begin{align}
\phi_{\UMP,\alpha}^n(k)=
\left\{
\begin{array}{ll}
1 & \hbox{ if } \frac{k}{n} <   \theta_0 
- 
\epsilon_\alpha \sqrt{\frac{\theta_0(1-\theta_0)}{n}} 
\\
0 & \hbox{ if } \frac{k}{n} \ge \theta_0 
- 
\epsilon_\alpha \sqrt{\frac{\theta_0(1-\theta_0)}{n}} .
\end{array}
\right.
\end{align}
The p-value is also approximated to 
\begin{align}
\Phi(\frac{k-n\theta_0}{\sqrt{n(1-\theta_0)\theta_0}}).
\label{3-6-1}
\end{align}

The Poisson distributions ${\rm Poi}(\mu)$
are also an exponential family
because another parameter 
$\theta:= \log \mu$ satisfies
${\rm Poi}(\mu)(n)=
\frac{1}{n\!}e^{\theta n -e^\theta}$.
The UMP test $\phi_{\UMP,\alpha}$ of the level $\alpha$
is characterized similarly to (\ref{5-4-1}).
When the threshold $\mu_0$ is sufficiently large
and 
the hypothesis is given 
\begin{equation}
	\label{eq:hypo-int-g}
	H_0:\mu\ge\mu_0
	\mbox{ versus }
	H_1:\mu < \mu_0,
\end{equation}
the UMP test $\phi_{\UMP,\alpha}$ of the level $\alpha$
is approximately given as
\begin{align}
\phi_{\UMP,\alpha}(n)=
\left\{
\begin{array}{ll}
1 & \hbox{ if } n <   \mu_0
- \epsilon_\alpha\sqrt{\mu_0} \\
0 & \hbox{ if } n \ge \mu_0 
- \epsilon_\alpha\sqrt{\mu_0}.
\end{array}
\right.
\end{align}
The p-value is also approximated to 
\begin{align}
\Phi(\frac{n-\mu_0}{\sqrt{\mu_0}}),
\label{3-6-3}
\end{align}

Next, we consider
testing 
the following hypothesis
in the case of the binomial Poisson distribution
Poi($\mu_1,\mu_2$):
\begin{equation}
	\label{eq:hypo-int-2}
	H_0:\frac{\mu_1}{\mu_1+\mu_2} \ge\theta_0
	\mbox{ versus }
	H_1:\frac{\mu_1}{\mu_1+\mu_2}  < \theta_0.
\end{equation}
In this case, as is shown at 
(\ref{5-14-1}) and (\ref{5-14-2})
in Section \ref{s4d},
the likelihood ratio test $\phi_{\LR,r}$ is 
characetrized by 
the likelihood ratio test of the binomial distributions
as
\begin{align}
\phi_{\LR,r}(n_1,n_2)=
\phi_{\LR,r}^{n_1+n_2}(n_1).
\end{align}
Hence, it is suitable to employ
the likelihood ratio test
$\phi_{\LR,l=\alpha}(n_1,n_2)=
\phi_{\UMP,\alpha}^{n_1+n_2}(n_1)$
with the level $\alpha$.
This is because the conditional distribution
$
\frac{{\rm Poi}(\mu_1,\mu_2)(n_1,n_2)}
{\sum_{k'=0}^{n_1+n_2}{\rm Poi}(\mu_1,\mu_2)(k',n_1+n_2-k')}
$ is equal to 
the binomial distribution
$P_{\frac{\mu_1}{\mu_1+\mu_2}}^{n_1+n_2}(n_1)$.
Therefore, when we observe $n_1,n_2$,
the p-value of this class of likelihood ratio tests
is equal to 
$\sum_{k=0}^{n_1}
\choose{n_1+n_2}{k}
\theta_0^k(1-\theta_0)^{n_1+n_2-k}$.

When the total number $n_1+n_2$ is sufficiently large,
the test $\phi_{\LR,l=\alpha}$ of the level $\alpha$
is approximately given as
\begin{align}
\phi_{\LR,l=\alpha}(n_1,n_2):=
\left\{
\begin{array}{ll}
1 & \hbox{ if } \frac{n_1}{n_1+n_2} 
<   \theta_0 
- 
\epsilon_\alpha
\sqrt{\frac{\theta_0(1-\theta_0)}{n_1+n_2}} 
\\
0 & \hbox{ if } \frac{n_1}{n_1+n_2}  \ge \theta_0 
- 
\epsilon_\alpha
\sqrt{\frac{\theta_0(1-\theta_0)}{n_1+n_2}} .
\end{array}
\right.
\end{align}
The p-value is also approximated to 
\begin{align}
\Phi(\frac{n_1-(n_1+n_2)\theta_0}{\sqrt{(n_1+n_2)(1-\theta_0)\theta_0}}).
\label{3-6-1-h}
\end{align}

\subsection{Multi-parameter case}\label{s4d}
In the one-parameter case, 
UMP tests can be often characterized by likelihood ratiotests.
However, in the multi-parameter case,
this type characterization is impossible generally,
and the UMP test does not always exist.
In this case, we have to choose our test among non-UMP tests.
One idea is choosing our test among likelihood ratio tests
because likelihood ratio tests always exist and 
we can expect that these tests have good performances.
Generally, 
it is not easy to give an explicit form of the likelihood ratio test.
When the family is 
a multi-parameter exponential family,
the likelihood ratio test has a simple form.
A family of probability distributions
$\{P_{\vec{\theta}}|\vec{\theta}=(\theta^1, \ldots, \theta^m) \in  \real^m\}$
is called an $m$-parameter exponential family
when there exists $m$-dimensional random variable $\vec{x}=(x_1, \ldots, x_m)$ 
such that
\[
	P_{\vec{\theta}}(\vec{x}):=P_0(\vec{x})
	\exp(\vec{\theta} \cdot \vec{x}+ g(\vec{\theta}))
	,
\]
where 
$g(\vec{\theta}):= -\log \int \exp(\vec{\theta} \cdot \vec{x})
P_0(d \vec{x})$.
However, this form is not sufficiently simple because
its rejection region is given by the a nonlinear constraint.
Hence, a test with a simpler form is required.
In the following, we discuss the likelihood ratio test in the case of 
multi-nomial Poisson distribution.
After this discussion, 
we propose an alternative test.

In an $m$-parameter exponential family,
the likelihood ratio test $\phi_{\LR,r}$ has the form
\begin{align}
\phi_{\LR,r}(\vec{x})=
\left\{
\begin{array}{ll}
0 \hbox{ if } &
\inf_{\vec{\theta}_1\in \Theta_1}D(P_{\vec{\theta}(\vec{x})}
\|P_{\vec{\theta}_1})  \\
&- \inf_{\vec{\theta}_0\in \Theta_0}
D(P_{\vec{\theta}(\vec{x})}\|P_{\vec{\theta}_0}) 
\ge \log r \\
1 \hbox{ if } & 
\inf_{\vec{\theta}_1\in \Theta_1}D(P_{\vec{\theta}(\vec{x})}
\|P_{\vec{\theta}_1}) \\
&- \inf_{\vec{\theta}_0\in \Theta_0}D(P_{\vec{\theta}(\vec{x})}
\|P_{\vec{\theta}_0}) 
< \log r ,
\end{array}
\right.\label{2-16-1}
\end{align}
where the divergence $D(P_{\vec{\eta}}\|P_{\vec{\theta}})$ is defined as
\begin{align*}
D(P_{\vec{\eta}}\|P_{\vec{\theta}})
&:=
\int \log \frac{P_{\vec{\eta}}(\vec{x}')}{P_{\vec{\theta}}(\vec{x}')}
P_{\vec{\eta}}(d \vec{x}')\\
&=
(\vec{\eta}-\vec{\theta})\int \vec{x} P_{\vec{\eta}}(d \vec{x})
+ g(\vec{\eta})-g(\vec{\theta}),
\end{align*}
and $\vec{\theta}(\vec{x})$ is defined by \cite{A-N}
\begin{align}
\int \vec{x}' P_{\vec{\theta}(\vec{x})}(d \vec{x}')=\vec{x}.
\end{align}
This is because the logarithm of the likelihood function is calculated
as
\begin{align*}
&\log 
\frac{\sup_{\vec{\theta}_0 \in \Theta_0} P_{\vec{\theta}_0}(\vec{x})}
{\sup_{\vec{\theta}_1 \in \Theta_1} P_{\vec{\theta}_1}(\vec{x})}\\
=&
\sup_{\vec{\theta}_0 \in \Theta_0}
\inf_{\vec{\theta}_1 \in \Theta_1}
\log \frac{P_{\vec{\theta}_0}(\vec{x})}{P_{\vec{\theta}_1}(\vec{x})}\\
=&\sup_{\vec{\theta}_0 \in \Theta_0}
\inf_{\vec{\theta}_1 \in \Theta_1}
(\vec{\theta}_0 - \vec{\theta}_1)\cdot \vec{x}
+g(\vec{\theta}_0)- g(\vec{\theta}_0) \\
=&\sup_{\vec{\theta}_0 \in \Theta_0}\inf_{\vec{\theta}_1 \in \Theta_1}
(\vec{\theta}_0 - \vec{\theta}_1)\cdot \int \vec{x}' 
P_{\vec{\theta}(\vec{x})}(d \vec{x}')
+g(\vec{\theta}_0)- g(\vec{\theta}_0) \\
=&\sup_{\vec{\theta}_0 \in \Theta_0}\inf_{\vec{\theta}_1 \in \Theta_1}
D(P_{\vec{\theta}(\vec{x})}\|P_{\vec{\theta}_1}) 
- D(P_{\vec{\theta}(\vec{x})}\|P_{\vec{\theta}_0}) \\
=& \inf_{\vec{\theta}_1\in \Theta_1}
D(P_{\vec{\theta}(\vec{x})}\|P_{\vec{\theta}_1}) 
- \inf_{\vec{\theta}_0\in \Theta_0}D(P_{\vec{\theta}(\vec{x})}
\|P_{\vec{\theta}_0}). 
\end{align*}
In addition, $\vec{\theta}(\vec{x})$ 
coincides with the MLE when $\vec{x}$ is observed.
Hence, when $\Theta= \Theta_0 \cup \Theta_1$,
the likelihood ratio test with the ratio $r<1$ is given by the rejection region:
\begin{align}
\left\{\vec{x}\left| \vec{\theta}(\vec{x})\in \Theta_1,
\inf_{\vec{\theta}_0\in \Theta_0 }
D(P_{\vec{\theta}(\vec{x})}\|P_{\vec{\theta}_0})
\ge  -\log r\right.\right\}.
\end{align}

In the case of the multi-nomial Poisson distributions 
Poi$(\vec{\mu})(\vec{k})
:= e^{-\sum_{i=1}^l \mu_i}
\frac{
\mu_1^{k_1} \cdots \mu_l^{k_m}}{k_1! \cdots k_m!}$,
which is an exponential family,
the divergence 
is calculated as
\begin{align}
&D({\rm Poi}(\vec{\mu})\|
{\rm Poi}(\vec{\mu}'))\nonumber\\
=&
\sum_{i=1}^m (\mu_i'- \mu_i)+ \sum_{i=1}^m \mu_i \log \frac{\mu_i}{\mu_i'}
\label{5-14-3}\\
=&
(\sum_{i=1}^m\mu_i')- (\sum_{i=1}^m\mu_i)
(\sum_{i=1}^m\mu_i)\log \frac{\sum_{i=1}^m\mu_i}{\sum_{i=1}^m\mu_i'}\nonumber \\
&+ 
(\sum_{i=1}^m\mu_i)
D(
\frac{\vec{\mu}}{\sum_{i=1}^m\mu_i}\|
\frac{\vec{\mu'}}{\sum_{i=1}^m\mu_i'}),\label{5-14-2}
\end{align}
where $D(\vec{p}\|\vec{p'})$ is the divergence between the multinomial 
distributions $\vec{p}$ and $\vec{p'}$.

When the hypothesis is given by (\ref{eq:hypo-int-2})
and 
$\frac{n_1}{n_1+n_2}\le \theta_0$,
we have
\begin{align}
&\log 
\frac{\sup_{\vec{\theta}_0 \in \Theta_0} P_{\vec{\theta}_0}
(n_1,n_2)
}
{\sup_{\vec{\theta}_1 \in \Theta_1} P_{\vec{\theta}_1}
(n_1,n_2)}\nonumber \\
=&
(n_1+n_2) D(P_{\frac{n_1}{n_1+n_2}}\| P_{\theta_0})
=
D(P_{\frac{n_1}{n_1+n_2}}^{n_1+n_2}
\| P_{\theta_0}^{n_1+n_2}),\label{5-14-1}
\end{align}
where $P_\theta$ is the binomial distribution with one observation
and $P_\theta^n$ is the binomial distribution with $n$ observations.
Then, the likelihood ratio test is given by the likelihood ratio test 
of the binomial distributions.

In the following, we treat
two hypotheses given as
\begin{equation}
	\label{eq:hypo-int-3}
	H_0:\vec{w} \cdot \vec{\theta}\ge c_0
	\mbox{ versus }
	H_1:\vec{w} \cdot \vec{\theta} < c_0,
\end{equation}
with the condition $w_i \ge 0$,
Using the formula (\ref{5-14-3}),
and (\ref{eq:hypo-int-3}), we can calculate
the likelihood ratio test for a given ratio $r$.
Now, we calculate 
the p-value concerning the class of likelihood ratio tests
when we observe the data $k_1,  \ldots, k_m$.
When $\vec{w} \cdot \vec{k}  < c_0$,
this p-value is equal to 
\begin{align}
\max_{\vec{w}\cdot \vec{\mu}' = c_0}
{\rm Poi}(
\vec{\mu'}
)
(\cA_{R(\vec{k})}),\label{3-7-7}
\end{align}
where
\begin{align*}
\cA_R&:=
\left\{
\vec{k'}
\left|
\begin{array}{l}
\displaystyle
\min_{\vec{w}\cdot \vec{\mu} = c_0}
\sum_{i=1}^m (\mu_i- k_i')+ \sum_{i=1}^m k_i' \log \frac{k_i'}{\mu_i} \ge 
R \\
\vec{w} \cdot \vec{k}'  < c_0
\end{array}
\right.\right\},\\
R(\vec{k})&:=
\min_{\vec{w}\cdot \vec{\mu} = c_0}
\sum_{i=1}^m (\mu_i- k_i)+ \sum_{i=1}^m k_i \log \frac{k_i}{\mu_i}
\end{align*}
because the minimum $R$ satisfying $\vec{k}\in \cA_R$ is $R(\vec{K})$.
Since the calculation of (\ref{3-7-7})
is not so easy,
we consider its upper bound.
For this purpose, we define 
the set $\cB_R$ as
\begin{align}
\cB_R :=
\left\{\vec{k}'\left|
\sum_{i=1}^m \frac{k_i'}{\tilde{\mu}_i(R)}
\le 1
\right.\right\},
\end{align}
where $\tilde{\mu}_i(R)$ are defined as follows:
\begin{align}
\frac{c_0}{w_i} - \tilde{\mu}_i(R)+ 
\tilde{\mu}_i(R)\log \frac{\tilde{\mu}_i(R) w_i}{c_0}&= R 
\hbox{~~if }
R\le R_{0,i}
\label{2-6-12}
\\
\frac{c_0}{w_M} + \tilde{\mu}_i(R) \log \frac{w_M-w_i}{w_M}&= R 
\hbox{~~if }
R > R_{0,i},
\label{2-6-11}
\end{align}
where 
$w_M:= \max_i w_i$
and 
$R_{0,i}:=\frac{c_0}{w_M} 
+ \frac{c_0(w_M-w_i)}{w_i w_M}\log \frac{w_M-w_i}{w_M}$.
Note that $\tilde{\mu}_i(R)$ is a monotone decreasing function of $R$.
As is shown in Appendix \ref{3-6-10}, 
\begin{align}
\cA_R \subset \cB_R \label{5-2-1}.
\end{align}
Then, the p-value concerning likelihood ratio tests 
is upperly bounded by 
\begin{align}
\max_{\vec{w}\cdot \vec{\mu}' = c_0}
{\rm Poi}(\vec{\mu}')(\cB_{R(\vec{k})}) \label{2-6-13}.
\end{align}

However, it is difficult to choose the likelihood 
$r$ such that the p-value is equal to a given risk probability $\alpha$
because the set $\cA_R$ is defined by a non-linear constraint.
In order to resolve this problem, we propose to modify
the likelihood ratio test by using the set $\cB_R$ 
instead of the set $\cA_R$
because $\cB_R$ is defined by a linear constraint
while $\cA_R$ is by a non-linear constraint.
That is, we define 
the modified test $\phi_{\new,R}$
as the test with the rejection region $\cB_{R}$.
Among this kind of tests,
we can choose the test $\phi_{\new,R_\alpha}$ 
with the risk probability $\alpha$ 
by choosing $R_{\alpha}$ in the following way:
\begin{align}
\max_{\vec{w}\cdot \vec{\mu}' = c_0}
{\rm Poi}(\vec{\mu}')(\cB_{R_\alpha})= \alpha.
\end{align}
Indeed, the calculation of the probability 
${\rm Poi}(\vec{\mu}')(\cB_{R})$ 
is easier than that of 
the probability 
${\rm Poi}(\vec{\mu}')(\cA_{R})$
because of the linearity of the constraint condition of $\cB_R$.

Next, we calculate 
the p-value of the set of the modified tests $\{\phi_{\new,\alpha}\}_{\alpha}$.
For an observed data $\vec{k}$, we choose 
$R'(\vec{k})$ as $R'$ satisfying 
\begin{align}
\sum_{i=1}^m \frac{k_i}{\tilde{\mu}_i(R')}=1.
\end{align}
The LHS is monotone increasing for $R'$
because each $\tilde{\mu}_i(R')$ is monotone decreasing for $R'$.
Thus, $R'(\vec{k})$ is the maximum $R'$ such that $\vec{k} \in \cB_{R'}$.
Then, the p-value is equal to 
$\max_{\vec{w}\cdot \vec{\mu}' = c_0}
{\rm Poi}(\vec{\mu}')(\cB_{R'(\vec{k})})$.
Further, the relation (\ref{5-2-1}) implies 
$\vec{k} \in \cB_{R(\vec{k})}$.
Hence, $R(\vec{k}) \le R'(\vec{k})$, which implies 
$\cB_{R(\vec{k})} \supset \cB_{R'(\vec{k})}$.
Therefore,
the p-value $\max_{\vec{w}\cdot \vec{\mu}' = c_0}
{\rm Poi}(\vec{\mu'})(\cB_{R'(\vec{k})})$
concerning the modified tests $\{\phi_{\new,\alpha}\}_{\alpha}$
is smaller than 
the upper bound 
$\max_{\vec{w}\cdot \vec{\mu}' = c_0}{\rm Poi}(\vec{\mu'})(\cB_{R(\vec{k})})$
of p-value concerning the likelihood ratio tests.
This test $\phi_{\new}$ coincides with the likelihood ratio test 
in the one-parameter case.


\section{Asymptotic Theory}\label{s5}
\subsection{Fisher information}\label{2-7-4}
Assume that the data $x_1, \ldots, x_n$ obeys 
the identical and independent distribution
of the same distribution family $p_\theta$
and $n$ is sufficiently large.
When the true parameter $\theta$ is close to $\theta_0$,
it is known that the meaningful information for $\theta$ is 
essentially given as the random variable
$\frac{1}{n}\sum_{i=1}^n l_{\theta_0}(x_i)$,
where the logarithmic derivative $l_{\theta_0}(x_i)$
is defined by
\begin{align}
l_{\theta}(x):= \frac{d \log p_\theta(x)}{d \theta}.
\end{align}
In this case, 
the random variable 
$\frac{1}{n} \sum_{i=1}^n l_{\theta_0}(x_i)$
can be approximated by the normal distribution
with the expectation value $\theta -\theta_0$
and the variance $\frac{1}{n J_{\theta_0}}$,
where the Fisher information $J_{\theta}$ is defined as
$J_{\theta}:=
\int (l_{\theta}(x))^2 P_\theta (d x)$.
Hence, 
the testing problem can be approximated by 
the testing of this normal distribution family \cite{A-N,lehmann}.
That is, the quality of testing is approximately evaluated by
the Fisher information $J_{\theta_0}$ at the threshold $\theta_0$.

In the case of Poisson distribution family 
Poi$(\theta t)$, 
the parameter $\theta$ can be estimated by $\frac{X}{t}$.
The asymptotic case corresponds to the case with large $t$.
In this case, Fisher information is $\frac{t}{\theta}$.
When $X$ obeys the unknown Poisson distribution family Poi$(\theta t)$, 
the estimation error
$\frac{X}{t}-\theta$ is close to the normal distribution
with the variance $\frac{\theta}{t}$,
{\it i.e.},
$\sqrt{t}(\frac{X}{t}-\theta)$ approaches to 
the random variables obeying the normal distribution with 
variance $\theta$.
That is, Fisher information corresponds to 
the inverse of variance of the estimator.

This approximation can be extended to the multi-parameter case
$\{p_{\theta}|\theta \in \real^m\}$.
Similarly, it is known that
the testing problem can be approximated by 
the testing of the normal distribution family
with the covariance matrix $(n J_{\theta})^{-1}$,
where the Fisher information matrix $J_{\theta;i,j}$
is given by
\begin{align}
J_{\theta;i,j}&:=
\int l_{\theta;i}(x)  l_{\theta;j}(x) P_\theta (d x),\\
l_{\theta;i}(x)&:= \frac{\partial \log p_\theta(x)}{\partial \theta^i}.
\end{align}
When the hypotheses is given by (\ref{eq:hypo-int}),
the testing problem can be approximated by 
the testing of the normal distribution family
with variance $\frac{\vec{w}\cdot J_{\theta_0}^{-1}\vec{w}}{n}$,

Indeed, the same fact holds for 
the multinomial Poisson distribution family 
Poi$(t \vec{\mu})$.
When the random variable $X_j$ is the $i$-th random variable,
the random variable $\sum_{j=1}^m \frac{\lambda_j}{\sqrt{t}} 
(X_j - \mu_j)$ converges to the random variable obeying the normal distribution
with the variance $\sum_{j=1}^m \lambda_j^2 \mu_j $
in distribution:
\begin{align}
\sum_{j=1}^m \frac{\lambda_j}{\sqrt{t}} 
(X_j - \mu_j) \stackrel{d}{\longrightarrow}\sum_{j=1}^m \lambda_j^2 \mu_j .
\label{3-7-1}
\end{align}
This convergence is compact uniform concerning the parameter 
$\vec{\mu}$.
In this case, the Fisher information matrix $J_\mu$ 
is the diagonal matrix with the diagonal elements
$(\frac{t}{\mu_1},\ldots, \frac{t}{\mu_m})$.
When our distribution family is given as a subfamily
Poi$(t \mu_1(\theta), \ldots, t \mu_m(\theta))$,
the Fisher information matrix is $\cA_\theta^t J_{\mu(\theta)}\cA_\theta$,
where $\cA_{\theta;i,j}=\frac{\partial \mu_j}{\partial \theta_i}$.
Hence, 
when the hypotheses is given by (\ref{eq:hypo-int-3}),
the testing problem can be approximated by 
the testing of the normal distribution family
with variance 
\begin{align}
\vec{w}\cdot (\cA_\theta^t J_{\mu(\theta)}\cA_\theta)^{-1} \vec{w}.
\label{2-16-5}
\end{align}
In the following, we call this value Fisher information.
Based on this value, 
the quality can be compared when we have several testing schemes.

\subsection{Multi-parametric Poisson distribution}
In the following, 
we treat testing of the hypothesis (\ref{eq:hypo-int-3}) in 
the multinomial Poisson distribution Poi($\vec{\mu}$)
by using normal approximation.
In this case,
by using $\tilde{\mu}_i$ defined in (\ref{2-6-12}) and (\ref{2-6-11}),
the upper bound (\ref{2-6-13})
of the p-value concerning
the likelihood ratio tests
is approximated to 
\begin{align*}
& \max_{w\cdot \mu' = c_0}
\Phi\left(
\frac{1- \sum_{j=1}^m
\frac{\mu_j'}{\tilde{\mu}_j(R(\vec{k}))}
}
{\sqrt{
\sum_{i=1}^m
\frac{\mu_i' }{\tilde{\mu}_i(R(\vec{k}))^2}
}}
\right)\\
=&
\Phi\left(
\max_{w\cdot \mu' = c_0}
\frac{1- \sum_{j=1}^m
\frac{\mu_j'}{\tilde{\mu}_j(R(\vec{k}))}
}
{\sqrt{
\sum_{i=1}^m
\frac{\mu_i' }{\tilde{\mu}_i(R(\vec{k}))^2}
}}
\right),
\end{align*}
because this convergence 
(\ref{3-7-1}) is compact uniform concerning the parameter 
$\vec{\mu}$.
Letting $x_i(R)= \frac{c_0}{w_i \tilde{\mu}_i(R)}-1$
and $y_i(R)= \frac{c_0}{w_i \tilde{\mu}_i(R)^2}$,
we have
\begin{align}
\max_{w\cdot \mu' = c_0}
\frac{1- \sum_{j=1}^m
\frac{\mu_j'}{\tilde{\mu}_j(R(\vec{k}))}}
{\sqrt{
\sum_{i=1}^m
\frac{\mu_i'}{\tilde{\mu}_i(R(\vec{k}))^2}
}}
= 
\max_{(x,y)\in Co(R(\vec{k}))} 
\frac{-x}{\sqrt{y}},
\end{align}
where 
$Co(R)$ is the convex hull of
$(x_1(R),y_1(R)), \ldots, (x_m(R),y_m(R))$.
As is shown in Appendix \ref{a5},
this value is simplified to
\begin{align}
- \min_{i,j}
z_{i,j}(R(\vec{k})),\label{5-5-1}
\end{align}
where
\begin{align}
z_{i,j}(R):=
\left\{
\begin{array}{ll}
\frac{x_i(R)}{\sqrt{y_i(R)}} & 
\hbox{ if }
\frac{2x_j(R)y_i(R)}{x_i(R)y_i(R)+x_i(R)y_j(R) }\ge 1 
\\
\frac{x_j(R)}{\sqrt{y_j(R)}} & 
\hbox{ if }
\frac{2x_i(R)y_j(R)}{x_j(R)y_j(R)+x_j(R)y_i(R) }\ge 1 
\\
\tilde{z}_{i,j}(R)
& \hbox{ otherwise},
\end{array}
\right.
\end{align}
where
\begin{widetext}
\begin{align}
\tilde{z}_{i,j}(R):=
\frac{2
(x_i(R) x_j(R)(y_i(R) +y_j(R))- x_i(R)^2 y_j(R)- x_j(R)^2 y_i(R))}{
\sqrt{(x_i(R)-x_j(R))(y_i(R)-y_j(R))}
\sqrt{x_i(R) y_j(R)^2+ x_j(R) y_i(R)^2 -y_i(R) y_j(R) (x_i(R)+x_j(R))}
}
\end{align}
\end{widetext}
That is, our upper bound of p-value concerning the likelihood ratio tests 
is given by 
\begin{align}
\displaystyle \Phi(
- \min_{i,j}
z_{i,j}(R(\vec{k}))
).\label{3-7-2}
\end{align}

Next, we approximately calculate the test with the risk probability
$\alpha$ proposed in section\ref{s4d}.
First, we choose $R_\alpha$ by
\begin{align}
- \min_{i,j}z_{i,j}(R_\alpha)= \Phi^{-1}(\alpha).
\end{align}
Then, our test is given by the rejection region $\cB_{R_\alpha}$.
Using the same discussion, the p-value concerning the proposed tests is
equal to 
\begin{align}
\displaystyle \Phi(
- \min_{i,j}
z_{i,j}(R'(\vec{k}))).\label{3-7-2-a}
\end{align}

\section{Modification of Visibility}\label{s6}
In the following sections,
we apply the discussions in sections \ref{s3} - \ref{s5}
to the hypothesis (\ref{5-5-2}).
That is, we consider how to reject the null hypothesis 
$H_0:\fid \le \fid_0$ with a certain risk probability $\alpha$.

In the usual visibility, we usually 
measure the coincidence events only in the one direction or two directions.
However, in this method, the number of the counts of coincidence events
be reflected not only by the fidelity but also by the direction of 
difference between the true state of target maximally entangled state.
In order to remove the bias based on such a direction,
we propose to measure 
the counts of 
the coincidence vectors
$|HH\rangle, |VV\rangle, |DD\rangle, |XX\rangle, 
|RL\rangle,$ and $|LR\rangle$,
which corresponds to the coincidence events,
and the counts of the anti-coincidence vectors
$|HV\rangle, |VH\rangle, |DX\rangle, |XD\rangle, 
|RR\rangle$, and $|LL\rangle$,
which corresponds to the anti-coincidence events.
The former corresponds to the 
the minimum values in the usual visibility,
and the later does to 
the minimum values in the usual visibility.
In this paper, we call this proposed method the modified visibility method.
Using this method, we can test the fidelity 
between the maximally entangled state 
$|\Phi^{(+)}\rangle \langle \Phi^{(+)}|$
and the given state $\sigma$, using
the total number of counts of the coincidence events 
(the total count on coincidence event)
$n_1$
and 
the total number of counts of the anti-coincidence events 
(the total count on anti-coincidence events)
$n_2$ obtained by measuring on all the vectors
with the time $\frac{t}{12}$.
When the dark count is negligible, 
the total count on coincidence events $n_1$ obeys 
Poi$(\lambda\frac{2\fid+1}{12}t)$,
and the count on total anti-coincidence events $n_2$
obeys the distribution Poi$(\lambda\frac{2-2\fid}{12}t)$.
These expectation values $\mu_{1}$ and $\mu_{2}$
are given as 
$\mu_{1} = \lambda\frac{2\fid+1}{12}t$ and 
$\mu_{2} = \lambda\frac{2-2\fid}{12}t$.
Hence, 
Fisher information matrix concerning the parameters $\fid$ and $\lambda$
is 
\begin{align}
\left(
\begin{array}{cc}
\lambda (\frac{t}{3(2\fid +1)}+\frac{t}{3(2-2\fid)})
&0 \\
0&
\frac{\frac{2\fid +1}{12}t + \frac{2-2\fid}{12}t}{\lambda}
\end{array}
\right),
\end{align}
where the first element corresponds to the parameter $\fid$ and 
the second one does to the parameter $\lambda$. 
Then, we can apply the test $\phi_{\LR}$ 
given in the end of subsection \ref{2-7-5}.
That is, 
based on the ratio
$\frac{\mu_{2}}{\mu_{1}+\mu_{2}}=\frac{2}{3}(1-\fid)$,
we estimate the fidelity using 
the ratio $\frac{n_2}{n_1+n_2}$ as
$\hat{\fid}(n_1,n_2)= 1- \frac{3}{2}\frac{n_2}{n_1+n_2}$.
Based on the discussion in subsection \ref{2-7-4},
its variance is asymptotically equal to 
\begin{align}
\frac{1}
{\lambda(\frac{t}{3(2\fid +1)}+\frac{t}{3(2-2\fid)})}
=
\frac{(2\fid +1)(2-2\fid)}{\lambda t}.
\end{align}
Hence, 
similarly to the visibility, 
we can check the fidelity by using this ratio.

Indeed, when we consider the distribution under the condition 
that the total count $n_1+n_2$ is fixed to $n$,
the random variable $n_2$ obeys the binomial distribution with
the average value $\frac{2}{3}(1-\fid)n$.
Hence, we can apply the likelihood ratio test 
of the binomial distribution.
In this case, by the approximation to the normal distribution,
the likelihood ratio test with the risk probability $\alpha$
is almost equal to the test with the rejection region:
$\{(n_1,n_2)|\frac{n_2}{n_1+n_2}\le 
\frac{2}{3}(1-\fid_0)+ 
\Phi^{-1}(\alpha)\sqrt{\frac{(2-2\fid_0)(1+2\fid_0)}{9(n_1+n_2)}}\}$
concerning the null hypothesis 
$H_0:\fid \le \fid_0$.
The p-value of this kind of tests is 
$\Phi(\frac{n_2(2\fid_{0}+1)- n_1 (2-2\fid_{0})}
{\sqrt{(n_1+n_2)(2\fid_{0}+1)(2-2\fid_{0})}})$.

\section{Design I ($\lambda$: unknown, One Stage)}\label{s7}
In this section, we consider 
the problem of testing the fidelity 
between the maximally entangled state 
$|\Phi^{(+)}\rangle \langle \Phi^{(+)}|$
and the given state $\sigma$
by performing three kinds of measurement,
coincidence, anti-coincidence, and total flux,
with the times $t_1,t_2$ and $t_3$, respectively.
When the dark count is negligible, 
the data $(n_1,n_2,n_3)$ obeys 
the multinomial Poisson distribution
Poi$(\lambda\frac{2\fid+1}{6}t_1,\lambda\frac{2-2\fid}{6}t_2,\lambda t_3)$
with the assumption that the parameter $\lambda$ is unknown.
In this problem, it is natural to assume that 
we can select the time allocation 
with the constraint for the total time $t_1+t_2+t_3= t$.

The performance of the time allocation 
$(t_1,t_2,t_3)$ can evaluated by the variance (\ref{2-16-5}).
The Fisher information matrix concerning the parameters
$\fid$ and $\lambda$ 
is 
\begin{align}
\left(
\begin{array}{cc}
\lambda (\frac{2 t_1}{3(2\fid +1)}+\frac{2 t_2}{3(2-2\fid)})
& \frac{t_1-t_2}{3} \\
\frac{t_1-t_2}{3} &
\frac{\frac{2\fid +1}{6}t_1 + \frac{2-2\fid}{6}t_2+t_3}{\lambda}
\end{array}
\right),
\end{align}
where the first element corresponds to the parameter $\fid$ and 
the second one does to the parameter $\lambda$. 
Then, the asymptotic variance (\ref{2-16-5}) is calculated as
\begin{align}
\frac{\frac{2\fid+1}{6}t_1 +\frac{2 -\fid}{6}t_2 + t_3 }
{\lambda\left(
(\frac{2\fid+1}{6}t_1 +\frac{2 -\fid}{6}t_2 + t_3 )
(\frac{2 t_1}{3(2\fid +1)}+\frac{2 t_2}{3(2-2\fid)})
- (\frac{t_1-t_2}{3})^2\right)}.\label{2-16-10}
\end{align}
We optimize the time allocation by minimizing the variance (\ref{2-16-10}).
We perform the minimization by maximizing the inverse:
$
\lambda \left(
\frac{2 t_1}{3(2\fid +1)}+\frac{2 t_2}{3(2-2\fid)}
- \frac{(\frac{t_1-t_2}{3})^2}
{\frac{2\fid+1}{6}t_1 +\frac{2-2\fid}{6}t_2 + t_3 }
\right)$.
Applying Lemmas \ref{2-24-6} and \ref{2-18-6} shown in Appendix \ref{app1}
to the case of
$a= \frac{2}{3(2\fid +1)}$,
$b=\frac{2}{3(2-2\fid)}$,
$c=\frac{2\fid+1}{6}$,
$d= \frac{2-2\fid}{6}$,
we obtain
\begin{align}
\intertext{(i) }
\lambda \max_{t_1+t_3=t}
\frac{2 t_1}{3(2\fid +1)}
- \frac{(\frac{t_1}{3})^2}{\frac{2\fid+1}{6}t_1 + t_3 }
=&
\frac{2\lambda t}{3(2\fid +1)(1+\sqrt{\frac{2\fid+1}{6}})^2}
\label{2-19-1}\\
\intertext{(ii) }
\lambda \max_{t_2+t_3=t}
\frac{2 t_2}{3(2-2\fid)}
- 
\frac{(\frac{t_2}{3})^2}
     {\frac{2-2\fid}{6}t_2 + t_3 }
=&
\frac{2 \lambda t}{3(2 -2\fid)(1+\sqrt{\frac{2-2\fid}{6}})^2}
\label{2-19-2}
\end{align}
and 
\begin{align}
\mbox{(iii) }
&\lambda \max_{t_1+t_2=t}
\frac{2t_1}{3(2\fid +1)}+\frac{2t_2}{3(2-2\fid)}
- \frac{(\frac{t_1-t_2}{3})^2}
{\frac{2\fid+1}{6}t_1 +\frac{2-2\fid}{6}t_2 }\nonumber \\
=&
\frac{\lambda
(\frac{1}{3}\sqrt{\frac{2-2\fid}{2\fid+1}}
+
\frac{1}{3}\sqrt{\frac{2\fid+1}{2-2\fid}})^2
t}{
(\sqrt{\frac{2\fid+1}{6}}+
\sqrt{\frac{2-2\fid}{6}})^2}\nonumber \\
=&
\frac{6\lambda t}
{(2\fid+1)(2-2\fid)(\sqrt{2\fid+1}+\sqrt{2-2\fid})^2}.
\label{2-19-3}
\end{align}
Then, these relations 
give the optimal 
time allocations between 
(i) coincidence and total flux measurements, 
(ii) anti-coincidence and total flux measurements,
and (iii) coincidence and anti-coincidence measurements, respectively. 
The ratio of (\ref{2-19-3}) to (\ref{2-19-1}) 
is equal to 
\begin{align}
\frac{3(\sqrt{6}+\sqrt{2\fid+1})^2}
{2(2-2\fid)(\sqrt{2\fid+1}+\sqrt{2-2\fid})^2}
>1\label{2-24-7},
\end{align}
as shown in Appendix \ref{2-24-10}. 
That is, the optimal 
measurement using the coincidence and the anti-coincidence 
always
provides 
better test than that using the coincidence and the total flux.
Hence, we compare (ii) with (iii), and obtain
\begin{align}
& \max_{t_1+t_2+t_3=t}
\lambda \Bigl(
\frac{2 t_1}{3(2\fid +1)}+\frac{2 t_2}{3(2-2\fid)}\nonumber\\
&\quad - \frac{(\frac{t_1-t_2}{3})^2}
{\frac{2\fid+1}{6}t_1 +\frac{2-2\fid}{6}t_2 + t_3 }
\Bigr)\nonumber \\
=&
\left\{
\begin{array}{ll}
\frac{4 \lambda t}{(2 -2\fid)(\sqrt{6}+\sqrt{2-2\fid})^2}& 
\hbox{ if } \fid_1 < \fid \le 1\\
\frac{6 \lambda t}{(2\fid+1)(2-2\fid)
(\sqrt{2\fid+1}+\sqrt{2-2\fid})^2}& 
\hbox{ if } 0 \le \fid \le \fid_1,
\end{array}
\right. \label{2-24-11}
\end{align}
where the critical point $\fid_1<1$ is defined by
\begin{align}
\frac{2(2\fid_1+1)(\sqrt{2\fid_1+1}+\sqrt{2-2\fid_1})^2}
{3(\sqrt{6}+\sqrt{2-2\fid_1})^2}
=1.
\end{align}
The approximated value of the critical point $\fid_1$ is $0.899519$.
The equation (\ref{2-24-11}) is derived in Appendix \ref{2-24-12}.

Fig. \ref{ratio1} shows 
the ratio of the optimal Fisher information 
based on the anti-coincidence and total flux measurements 
to that based on the coincidence and anti-coincidence measurements. 
When $\fid_1 \le \fid \le 1$,
the maximum Fisher information is attained by 
$t_1=0$, $t_2= \frac{\sqrt{6}}{(\sqrt{6}+\sqrt{2(1-\fid)})}t$,
$t_3 = \frac{\sqrt{2(1-\fid)}}{\sqrt{6}+\sqrt{2(1-\fid)}}t$.
Otherwise, 
the maximum is attained by 
$t_1= \frac{\sqrt{2-2\fid}}{\sqrt{2\fid+1}+\sqrt{2-2\fid}}t$,
$t_2= \frac{\sqrt{2\fid+1}}{\sqrt{2\fid+1}+\sqrt{2-2\fid}}t$,
$t_3=0$.
The optimal time allocation shown in Fig. \ref{ratio1} implies
that we should measure the counts on the anti-coincidence vectors
preferentially 
over other vectors.

\begin{figure}[htbp]
     \begin{center}
  \includegraphics*[width=8cm]{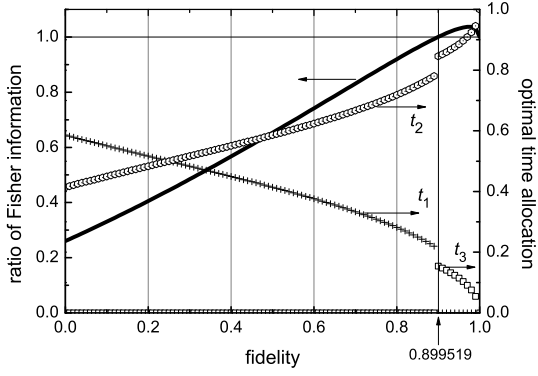}
  \end{center}
\caption{The ratio of the optimal Fisher information (solid line) and the optimal time allocation as a function of the fidelity $\fid$. The measurement time is divided 
into three periods: coincidence $t_{1}$ (plus signs), anti-coincidence $t_{2}$ (circles), and total flux $t_{3}$ (squares), which are normalized as $t_{1}+t_{2}+t_{3}=1$ in the plot.}
   \label{ratio1}
\end{figure}

The optimal asymptotic variance is 
$\frac{(2\fid+1)(2-2\fid)(\sqrt{2-2\fid}+\sqrt{1+2\fid})^2}
{6\lambda t}$ when the threshold $\fid_0$ is less than the critical point 
$\fid_1$.
This asymptotic variance is much better than that obtained 
by the modified visibility method.
The ratio of the optimal asymptotic variance is given by
\begin{align}
\frac{(\sqrt{2-2\fid}+\sqrt{1+2\fid})^2}{6} <1.
\end{align}

In the following, we give 
the optimal test of level $\alpha$ in the hypothesis testing 
(\ref{eq:hypo}).
Assume that the threshold $\fid_0$ is less than the critical point $\fid_1$.
In this case, we can apply testing of the hypothesis (\ref{eq:hypo-int-2}).
First, we measure the count on the coincidence vectors 
for a period of $t_1= \frac{t\sqrt{2-2\fid_0}}{\sqrt{2\fid_0+1}+\sqrt{2-2\fid_0}}$,
to obtain the total count $n_{1}$. 
Then, we measure the count on the anti-coincidence
vectors for a period of 
$t_2= \frac{t\sqrt{2\fid_0+1}}{\sqrt{2\fid_0+1}+\sqrt{2-2\fid_0}}$ 
to obtain the total count $n_{2}$.
Note that the optimal time allocation depends on the threshold of our 
hypothesis.
Finally, we apply the UMP test of $\alpha$ of
the hypothesis:
\begin{align*}
\begin{array}{ccc}
H_0: p \ge \frac{\sqrt{2-2\fid_0}}{\sqrt{2-2\fid_0}+\sqrt{1+2\fid_0}}
&\hbox{ versus }&
H_1: p < \frac{\sqrt{2-2\fid_0}}{\sqrt{2-2\fid_0}+\sqrt{1+2\fid_0}}
\end{array}
\end{align*}
with the binomial distribution family $P_{p}^{n_1+n_2}$
to the data $n_1$.
In this case, 
the likelihood ratio test with the risk probability $\alpha$
is almost equal to the test with the rejection region:
$\{(n_1,n_2)|
\frac{n_2}{n_2+n_1} \le 
\frac{\sqrt{2-2\fid_0}}{\sqrt{2\fid_0+1}+\sqrt{2-2\fid_0}}
+ \frac{\Phi^{-1}(\alpha)}{\sqrt{2\fid_0+1}+\sqrt{2-2\fid_0}}
\sqrt{\frac{\sqrt{2-2\fid_0}\sqrt{2\fid_0+1}}
{n_1+n_2}}\}$
concerning the null hypothesis 
$H_0:\fid \le \fid_0$.
The p-value of this kind of tests is 
$\Phi(
\frac{n_2 \sqrt{2\fid_{0}+1}- n_1 \sqrt{2-2\fid_{0}}}
{\sqrt{(n_1+n_2)\sqrt{2\fid_{0}+1}\sqrt{2-2\fid_{0}}}})$.

We can apply a similar testing for $\fid_0 > \fid_1$.
It is sufficient to replace the time allocation to
$t_1= 0$
$t_2 = \frac{t\sqrt{6}}{\sqrt{6}+\sqrt{2(1-\fid_0)}}$,
$t_3 = \frac{t\sqrt{2(1-\fid_0)}}{\sqrt{6}+\sqrt{2(1-\fid_0)}}$.
In this case, 
the likelihood ratio test with the risk probability $\alpha$
is almost equal to the test with the rejection region:
$\{(n_2,n_3)|
\frac{n_2}{n_2+n_3} \le 
\frac{\sqrt{1-\fid_0}}{\sqrt{3}+\sqrt{1-\fid_0}}
+ \frac{\Phi^{-1}(\alpha)}{\sqrt{3}+\sqrt{1-\fid_0}}
\sqrt{\frac{\sqrt{1-\fid_0}\sqrt{3}}
{n_2+n_3}}\}$
concerning the null hypothesis 
$H_0:\fid \le \fid_0$.
The p-value of this kind of tests is 
$\Phi(
\frac{n_2 \sqrt{3} - n_3 \sqrt{1-\fid_0}}
{\sqrt{(n_2+n_3) \sqrt{1-\fid_0}\sqrt{3}}})$.

Next, we consider the case where
the dark count parameter $\delta$ 
is known but is not negligible,
the Fisher information matrix is given by
\begin{widetext}
\begin{align}
\left(
\begin{array}{cc}
\lambda 
(\frac{2\lambda t_1}
{3(\lambda (2\fid +1)+6 \delta)}
+\frac{2\lambda t_2}
{3(\lambda (2-2\fid) +6 \delta) })
& 
\frac{\lambda (2\fid+1)}{3(\lambda (2\fid+1)+6 \delta)}t_1
-
\frac{\lambda (2-2\fid)}{3(\lambda (2-2\fid)+6 \delta)}t_2\\
\frac{\lambda (2\fid+1)}{3(\lambda (2\fid+1)+6 \delta)}t_1
-
\frac{\lambda (2-2\fid)}{3(\lambda (2-2\fid)+6 \delta)}t_2&
\frac{2\fid+1}{\lambda (2\fid+1)+6 \delta}
\frac{2\fid +1}{6}t_1 
+ 
\frac{2-2\fid}{\lambda (2-2\fid)+6 \delta}
\frac{2-2\fid}{6}t_2
+\frac{1}{\lambda}t_3
\end{array}
\right).
\end{align}
\end{widetext}
Hence, from (\ref{2-16-5}), the inverse of the minimum variance is 
equal to
\begin{align*}
& f(t_1,t_2,t_3)\\
:= &
\lambda 
(
\frac{2 \lambda t_1}{3(\lambda (2\fid +1)+6 \delta)}
+\frac{2 \lambda t_2}{3(\lambda (2-2\fid) +6 \delta) }\\
&
-\frac{(\frac{\lambda (2\fid+1)}{3(\lambda (2\fid+1)+6 \delta)}t_1
-
\frac{\lambda (2-2\fid)}{3(\lambda (2-2\fid)+6 \delta)}t_2)^2}
{
\frac{\lambda (2\fid+1)}{\lambda (2\fid+1)+6 \delta}
\frac{2\fid +1}{6}t_1 
+ 
\frac{\lambda (2-2\fid)}{\lambda (2-2\fid)+6 \delta}
\frac{2-2\fid}{6}t_2
+t_3}).
\end{align*}
Then, we apply Lemmas \ref{2-24-6} and \ref{2-18-6} in Appendix \ref{app1} to 
$\frac{f(t_1,t_2,t_3)}{\lambda}$
with $a=\frac{2\lambda}{3(\lambda (2\fid +1)+6 \delta)}$,
$b=\frac{2\lambda}{3(\lambda (2-2\fid) +6 \delta)}$,
$c=
\frac{\lambda (2\fid+1)}{\lambda (2\fid+1)+6 \delta}
\frac{2\fid +1}{6}$,
$d=\frac{\lambda (2-2\fid)}{\lambda (2-2\fid)+6 \delta}
\frac{2-2\fid}{6}$,
and obtain the optimized value:
\begin{align}
\intertext{(i) coincidence and total flux}
\max_{t_1+t_3=t}f(t_1,0,t_3)
&=
\frac{4\lambda t}{
((2\fid+1)+
\sqrt{\frac{6(\lambda (2\fid +1)+6 \delta)}{\lambda}})^2}
\label{2-24-14}
\\
\intertext{(ii) anti-coincidence and total flux}
\max_{t_2+t_3=t}f(0,t_2,t_3)
&=
\frac{4\lambda t}{
((2-2\fid)+
\sqrt{\frac{6(\lambda (2-2\fid )+6 \delta)}{\lambda}})^2}
\label{2-24-15}
\end{align}
and 
\begin{widetext}
\begin{align}
\intertext{(iii) coincidence and anti-coincidence}
 \max_{t_1+t_2=t}f(t_1,t_2,0) 
=&
\lambda t 
\left(\frac{
\frac{\lambda(2\fid+1)}
{3\sqrt{(\lambda(2\fid+1)+6 \delta)(\lambda(2-2\fid)+6\delta)}
}
+
\frac{\lambda(2-2\fid)}
{3\sqrt{(\lambda(2\fid+1)+6 \delta)(\lambda(2-2\fid)+6\delta)}
}
}
{
(2\fid+1)\sqrt{\frac{\lambda}{6(\lambda(2\fid+1)+6\delta)}}
+
(2-2\fid)\sqrt{\frac{\lambda}{6(\lambda(2-2\fid)+6\delta)}}
}\right)^2\nonumber \\
=&
\frac{2\lambda^2 t 
}{3(\lambda(2\fid+1)+6 \delta)(\lambda(2-2\fid)+6\delta)}
\left(\frac{3}
{
\frac{2\fid+1}{\sqrt{\lambda(2\fid+1)+6\delta}}
+
\frac{2-2\fid}{\sqrt{\lambda(2-2\fid)+6\delta}}
}\right)^2\nonumber \\
=&
\frac{6\lambda^2 t 
}{
\left(
(2\fid+1)
\sqrt{\lambda(2-2\fid)+6\delta}
+
(2-2\fid)
\sqrt{\lambda(2\fid+1)+6\delta}
\right)^2}.\label{2-24-16}
\end{align}
The ratio of (\ref{2-24-14}) to (\ref{2-24-16}) is 
\begin{align}
&\frac{3\lambda 
\left((2\fid+1)+
\sqrt{\frac{6(\lambda (2\fid +1)+6 \delta)}{\lambda}}\right)^2}
{2
\left(
(2\fid+1)
\sqrt{\lambda(2-2\fid)+6\delta}
+
(2-2\fid)
\sqrt{\lambda(2\fid+1)+6\delta}
\right)^2}\nonumber\\
=&
\frac{3}{2}
\left(
\frac{
(2\fid+1)\sqrt{\lambda}+
\sqrt{6(\lambda (2\fid +1)+6 \delta)}
}
{
(2\fid+1)
\sqrt{\lambda(2-2\fid)+6\delta}
+
(2-2\fid)
\sqrt{\lambda(2\fid+1)+6\delta}
}\right)^2 >1,\label{2-24-1-a}
\end{align}
\end{widetext}
where the final inequality is derived in Appendix \ref{2-24-10}. 
Therefore, the measurement using the coincidence and the anti-coincidence provides 
better test than that using the coincidence and the total flux, 
as in the case of $\delta=0$. 

\begin{figure}[htbp]
\vskip2cm
     \begin{center}
 \includegraphics*[width=8cm]{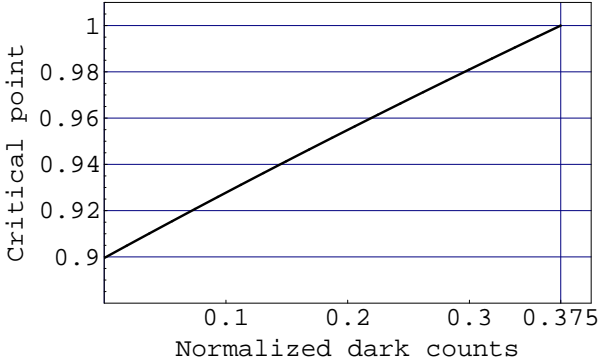}
  \end{center}
    \caption{The critical point  $\fid_{\delta'}$ for optimal time allocation 
as a function of normalized dark counts $\delta'$.} \label{thresh1}
\end{figure}

Define $\delta_1$ and the critical point $\fid_{\delta'}$ for 
the normalized dark count
$\delta' = 6 \delta/\lambda < \delta_1$
as
\begin{align*}
\sqrt{\delta_1+3}- \sqrt{\delta_1}&= \sqrt{3/2}\\
\sqrt{1+2\fid_{\delta'}+{\delta'}}-\sqrt{2-2\fid_{\delta'}+{\delta'}}
&= \sqrt{3/2}.
\end{align*}
The parameter $\delta_1$ is calculated to be $0.375$.
As shown in Appendix \ref{2-24-12}, the measurement using the coincidence 
and the anti-coincidence provides better test than 
that using the anti-coincidence and the total flux, if the fidelity is smaller than 
the critical point $\fid_{\delta'}$:
\begin{align}
&\max_{t_1+t_2+t_3=t}f(t_1,t_2,t_3) \nonumber \\
=&
\left\{
\begin{array}{ll}
\frac{4 \lambda^2 t}{
((2-2\fid)\sqrt{\lambda}+
\sqrt{6(\lambda (2-2\fid )+6 \delta)})^2}& 
\hbox{ if }
\fid > \fid_{\delta'} \\
\frac{6\lambda^2 t 
}{
\left(
(2\fid+1)
\sqrt{\lambda(2-2\fid)+6\delta}
+
(2-2\fid)
\sqrt{\lambda(2\fid+1)+6\delta}
\right)^2} & \hbox{ otherwise}.
\end{array}
\right. \label{2-24-3}
\end{align}
The optimal time allocation is given by
$t_1=0$, $t_2= 
\frac{t
\sqrt{6(\lambda(2-2\fid)+6\delta)}
}
{
\sqrt{6(\lambda(2-2\fid)+6\delta)}
+ (2-2\fid)\sqrt{\lambda}
}
$, and $t_3
=
\frac{t
(2-2\fid)\sqrt{\lambda}
}
{
\sqrt{6(\lambda(2-2\fid)+6\delta)}
+ (2-2\fid)\sqrt{\lambda}
}
$ for $\fid > \fid_{\delta'}$, and
$t_1=
\frac{t
(2-2\fid)\sqrt{\lambda (2\fid+1)+6\delta}
}
{
(2-2\fid)\sqrt{\lambda (2\fid+1)+6\delta}
+
(2\fid+1)\sqrt{\lambda (2-2\fid)+6\delta}
}$,
$t_2=
\frac{t
(2\fid+1)\sqrt{\lambda (2-2\fid)+6\delta}
}
{
(2-2\fid)\sqrt{\lambda (2\fid+1)+6\delta}
+
(2\fid+1)\sqrt{\lambda (2-2\fid)+6\delta}
}$,
$t_3=0$
for $\fid \le \fid_{\delta'}$.
The critical point  $\fid_{\delta'}$ for optimal time allocation 
increases with the normalized dark count
as illustrated in Fig. \ref{thresh1}. 

\section{Design II ($\lambda$: known, One Stage)}\label{s8}
In this section, we consider the case where $\lambda$ is known.
Then, the Fisher information is
\begin{align}
\lambda 
(
\frac{2\lambda t_1}
{3(\lambda (2\fid +1)+6 \delta)}
+\frac{2\lambda t_2}{3
(\lambda (2-2\fid) +6 \delta) }
).\label{2-24-1}
\end{align}
The maximum value is calculated as
\begin{align}
\max_{t_1+t_2+t_3=t}(\ref{2-24-1})
=
\left\{
\begin{array}{ll}
\frac{2\lambda^2 t}
{3(\lambda (2\fid +1)+6 \delta)}
& \hbox { if } \fid < \frac{1}{4} \\
\frac{2\lambda^2 t}{3
(\lambda (2-2\fid) +6 \delta) }
& \hbox { if } \fid \ge \frac{1}{4} .
\end{array}
\right.
\end{align}
The above optimization shows that when $\fid \ge \frac{1}{4} $,
the count on anti-coincidence 
$(t_1 = 0;t_2 =t;t_3=0)$ is better than
the count on coincidence 
$(t_1 = t;t_2 = 0;t_3=0)$.
In fact, Barbieri {\it et al.}\cite{BMNMDM03}
measured the sum of 
the counts on the anti-coincidence vectors
$|HV\rangle, |VH\rangle, |DX\rangle, |XD\rangle, |RR\rangle, |LL\rangle$
to realize the entanglement witness in their experiment.
In this case, the variance is
$\frac{3(\lambda (2-2\fid) +6 \delta) }{2\lambda^2 t}$.
When we observe the sum of counts on anti-coincidence $n_2$,
the estimated value of $\fid$ is given by 
$1+3(\delta-\frac{n_2}{\lambda t}) $,
which is the solution of 
$(\lambda \frac{2-2\fid}{6}+ \delta)t= n_2 $.
The likelihood ratio test with the risk probability $\alpha$
can be approximated by the test with the rejection region:
$\{n_2|n_2 \le
(\frac{\lambda (1-\fid_0)}{3}+\delta)t
+ \Phi^{-1}(\alpha)
\sqrt{
(\frac{\lambda (1-\fid_0)}{3}+\delta)t
}\}$
concerning the null hypothesis 
$H_0:\fid \le \fid_0$,
which is also the UMP test.
The p-value of likelihood ratio tests 
is 
$\Phi(\frac{n_2- (\frac{\lambda (1-\fid_0)}{3}+\delta)t
}{\sqrt{(\frac{\lambda (1-\fid_0)}{3}+\delta)t}})$.

When $\fid < \frac{1}{4}$,
the optimal time allocation is 
$t_1=t$, $t_2=t_3=0$.
The fidelity is estimated by $\frac{3 n_2}{\lambda t}-\frac{1}{2}$.
Its variance is 
$\frac{3(\lambda (2\fid +1)+6 \delta)}{2\lambda^2 t}$.
The likelihood ratio test with the risk probability $\alpha$
of the Poisson distribution
is almost equal to the test with the rejection region:
$\{n_1|n_1\ge
(\lambda \frac{1+2\fid_0}{6}+\delta)t
+ \Phi^{-1}(1-\alpha)
\sqrt{
(\lambda \frac{1+2\fid_0}{6}+\delta)t
}\}$
concerning the null hypothesis 
$H_0:\fid \le \fid_0$,
which is also the UMP test.
The p-value of likelihood ratio tests is
$\Phi(
\frac{-n_1+(\lambda \frac{1+2\fid_0}{6}+\delta)t}
{\sqrt{(\lambda \frac{1+2\fid_0}{6}+\delta)t}})$.

\section{Comparison of the asymptotic variances}\label{s9}
\begin{figure}[htbp]
\vskip2cm
     \begin{center}
 \includegraphics*[width=8cm]{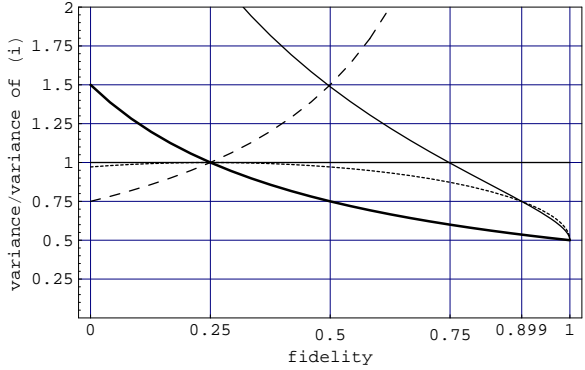}
  \end{center}
   \caption{
Comparison of the designs for time allocation. The asymptotic variances 
normalized by the value of modified visibility method are shown as a function of fidelity,
where dots: (iia), solid: (iib), thick: (iiia), and dash: (iiib). 
}
   \label{fig:relent9}
\end{figure}

We compare the asymptotic variances
of the following designs for time allocation, when the dark count $\delta$ parameter is zero.
\begin{description}
\item[(i)] 
 Modified visibility:
The asymptotic variance is 
$\frac{(2\fid +1)(2-2\fid)}{\lambda t}$.
\item[(iia)]
 Design I ($\lambda$ unknown).  optimal time allocation  between
the counts on anti-coincidence 
and coincidence:
The asymptotic variance is $\frac{(2\fid+1)(2-2\fid)
(\sqrt{2\fid+1}+\sqrt{2-2\fid})^2}{6 \lambda t}$.
\item[(iib)]
 Design I ($\lambda$ unknown), optimal time allocation between 
the counts on anti-coincidence and the total flux:
The asymptotic variance is $\frac{(2 -2\fid)(\sqrt{6}+\sqrt{2-2\fid})^2}{4 \lambda t}$.
\item[(iiia)]
 Design II  ($\lambda$ known), estimation from the count on anti-coincidence:
The asymptotic variance is $\frac{3(2-2\fid )}{2\lambda t}$.
\item[(iiib)]
 Design II  ($\lambda$ known), estimation from the count on coincidence:
The asymptotic variance is $\frac{3(2\fid +1)}{2\lambda t}$.
\end{description}
Fig. \ref{fig:relent9} shows the comparison, where the asymptotic variances 
in (iia)-(iiib) are normalized by the one in (i). The anti-coincidence 
measurement provides the best estimation for high ($\fid>0.25$) fidelity. 
When $\lambda$ is unknown, the measurement with the counts on anti-coincidence 
and the coincidence is better than that with the counts 
anti-coincidence and the total flux for $\fid < 0.899519$. 
For higher fidelity, the counts on anti-coincidence 
and total flux turns to be better, but the difference is small.

\section{Design III ($\lambda$: known, Two Stage)}\label{s10}

\subsection{Optimal Allocation}
The comparison in the previous section shows 
that the measurement on the anti-coincidence vectors
yields a better variance than 
the measurement on the coincidence vectors, 
when the fidelity is greater than $1/4$
and the parameters $\lambda$ and $\delta$ are known.
We will explore further improvement in the measurement on 
the anti-coincidence vectors.
In the previous sections, 
we allocate an equal time to the measurement on each of 
the anti-coincidence vectors.
Here we minimize the variance by optimizing the time allocation 
$t_{HV}$, $t_{VH}$, $t_{DX}$, $t_{XD}$, $t_{RR}$, and $t_{LL}$ between the 
anti-coincidence vectors
$B=\{|HV\rangle$, $|VH\rangle$, $|DX\rangle$, $|XD\rangle$, 
$|RR\rangle$, and $|LL\rangle\}$, 
under the restriction of the total measurement time:
$\sum_{(x,y)\in B} t_{x,y}=t$.
The number of the counts $n_{xy}$ obeys Poisson distribution
Poi($(\lambda \mu_{xy}+\delta)t_{xy}$) with unknown parameter $\mu_{xy}$.
Then, the Fisher information matrix is 
the diagonal matrix with the diagonal elements
$\{\frac{\lambda^2 t_{x,y}}{\lambda\mu_{x,y}+\delta}\}_{(x,y)\in B}$
Since we are interested in the parameter
$1-\fid=
\frac{1}{2}(\sum_{(x,y)\in B}
\mu_{x,y})$,
the variance is given by
\begin{align}
\frac{1}{4}
\Bigl(\sum_{(x,y)\in B}
\frac{\lambda\mu_{x,y}+\delta}{\lambda^2 t_{x,y}}
\Bigr), \label{2-24-30}
\end{align}
as mentioned in section \ref{2-7-4}.
Under the restriction of the total measurement time,
the minimum value of (\ref{2-24-30}) is 
\begin{align}
\frac{(
\sum_{(x,y)\in B} 
\sqrt{\lambda\mu_{x,y}+\delta}
)^2}{4 \lambda^2 t},\label{2-25-11}
\end{align}
which is attained by the optimal time allocation
\begin{align}
	t_{x y}=
	\frac
	{(\lambda\sqrt{\mu_{x y}}+\delta)t}
	{
\sum_{(x',y')\in B} \sqrt{\lambda\mu_{x',y'}+\delta}
},
\end{align}
which is called Neyman allocation and is used in sampling design\cite{Cochran}.
The variance with the equal allocation is
\begin{align}
\frac{3(\lambda (2-2\fid) +6 \delta) }{2\lambda^2 t}
=
\frac{3(\lambda (
\sum_{(x,y)\in B} 
\mu_{x,y}
) +6 \delta) }{2\lambda^2 t}. \label{2-25-10}
\end{align}
The inequality (\ref{2-25-11}) $\le$ (\ref{2-25-10}) 
can be derived from Schwartz's inequality of the vectors
$(1,\ldots,1)$ and
$(\sqrt{\lambda\mu_{HV}+\delta},
\ldots, \sqrt{\lambda\mu_{LL}+\delta})$.
The equality holds if and only if
$\mu_{HV}=\mu_{VH}=\mu_{DX}=\mu_{XD}=\mu_{RR}=\mu_{LL}$.
Therefore, the Neyman allocation has an advantage over the equal allocation,
when there is a bias in the parameters
$\mu_{HV},\mu_{VH},\mu_{DX},\mu_{XD},\mu_{RR},\mu_{LL}$. 
In other words, the Neyman allocation is effective 
when the expectation values of the counts on some vectors
are larger than
those on other vectors.

\subsection{Two-stage Method}
The optimal time allocation derived above 
is not applicable in the experiment, because it depends on the unknown parameters
$\mu_{HV},$ $\mu_{VH},$ $\mu_{DX},$ $\mu_{XD},$ $\mu_{RR},$ and $\mu_{LL}$.
In order to resolve this problem, 
we introduce a two-stage method, where 
the total measurement time $t$ is divided into
$t_f$ for the first stage and
$t_s$ for the second stage under the condition of $t=t_f+t_s$.
In the first stage,
we measure the counts on each vectors for $t_f/6$ and
estimate the expectation value for Neyman allocation on measurement time $t_s$.
In the second stage,
we measure the counts on a vector $\ket{x_A y_B}$ according to
the estimated Neyman allocation.
The two-stage method is formulated as follows.
\smallskip\\
(i)  The measurement time for each vector
in the first stage is given by
$t_f/6$ 
\\
(ii)
In the second stage,
we measure the counts on a vector $\ket{x_A y_B}$ 
with the measurement time $\tilde t_{x y}$
defined as
\[
	\tilde t_{x y}=
	\frac
	{m_{x y}}
	{\sum_{(x,y)\in B}\sqrt{m_{x y}}}
	(t-t_f)
\]
where
$m_{x y}$ is the observed count in the first stage.
\\
(iii)
Define $\hat\mu_{x y}$ and $\hat\fid$ as
\begin{align*}
	\hat{\mu}_{x y}=
	\frac{n_{x y}}{\lambda\tilde t_{x y}},
\quad	\hat\fid=
	1-\frac12\sum_{(x,y)\in B}\hat{\mu}_{x,y},
\end{align*}
where
$n_{x,y}$ is the number of the counts on $\ket{x_A y_B}$
for $\tilde t_{x y}$.
Then, we can estimate the fidelity by $\hat{\fid}$.
\\
(iv) Finally, we apply 
the test $\phi_{\new,\alpha}$ given in Section \ref{s4d}
to the two hypotheses given as
\begin{equation}
	H_0:\vec{w} \cdot \vec{\mu}\ge c_0
	\mbox{ versus }
	H_1:\vec{w} \cdot \vec{\mu} < c_0,
\end{equation}
where
$w_{x,y}:= \frac{1}{2\lambda \tilde{t}_{x,y}}$
and $c_0:= 1- \fid_0$.

\section{Conclusion}\label{s11}
We have formulated the hypothesis testing scheme to test the entanglement 
in the Poisson distribution framework.
Our statistical method can handle the fluctuation in the experimental data
more properly in a realistic setting. 
It has been shown that the optimal time allocation improves the test:
the measurement time should be allocated preferably to the 
anti-coincidence vectors.
This test is valid even if the dark count exists.
This design is particularly useful for the experimental test, because the
optimal time allocation depends only on the threshold of the test.
We don't need any further information of the probability distribution and 
the tested state.
The test can be further improved by optimizing time allocation 
between the anti-coincidence vectors, 
when the error from the maximally 
entangled state is anisotropic. However, this time allocation requires 
the expectation values on the  counts on coincidence, 
so that we need to apply the two stage method.

\section*{Acknowledgments}
The authors would like to thank Professor Hiroshi Imai of the
ERATO-SORST, QCI project for support.
They are grateful to Dr. Tohya Hiroshima, Dr. Yoshiyuki Tsuda
for useful discussions.

\appendix
\section{Optimization of Fisher information} \label{app1}
In this section, we maximize 
the quantities appearing in Fisher information.
\begin{lem}\label{2-24-6}
The equation 
\begin{align}
\max_{t_1,t_3 \ge 0,~c t_1+t_3= t}a t_2-\frac{a c t_1^2}{c t_1 +t_3}
=
\frac{at}{(\sqrt{c}+1)^2}\label{2-18-1}
\end{align}
holds and the maximum value is attained when
$t_1=\frac{t }{\sqrt{c}+1}$,
$t_3=\frac{\sqrt{c} t}{\sqrt{c}+1}$.
\end{lem}
\begin{proof}
Letting $x:=c t_1 +t_3$, we have
$t_1= \frac{x-t}{c-1}$.
Then, 
\begin{align*}
a t_1
-\frac{a c t_1^2}{c t_1 +t_3}
=\frac{a}{(c-1)^2}\left(-x -\frac{c t^2}{x}+(c+1)t\right).
\end{align*}
Hence, the maximum is attained at $x= \sqrt{c}t$,
{\it i.e.,}
$t_1= \frac{t}{\sqrt{c}+1}$ and 
$t_3= \frac{t\sqrt{c}}{\sqrt{c}+1}$.
Thus,
\begin{align*}
&\max_{t_1,t_3 \ge 0,~c t_1+t_3= t}
a t_1
-\frac{a c t_1^2}{c t_1 +t_3} \\
=&\frac{a}{(c-1)^2}\left(
-2\sqrt{c} t+(c+1)t\right)
=
\frac{at}{(\sqrt{c}+1)^2}.
\end{align*}
\end{proof}

\begin{lem}\label{2-18-6}
The equation
\begin{align}
&\max_{t_1,t_2 \ge 0,~t_1+t_2= t}
a t_1+ b t_2
-\frac{( \sqrt{ac}t_1-\sqrt{bc}t_2)^2}{c t_1 +d t_2 }\nonumber \\
=&
\frac{t(\sqrt{ad}+\sqrt{bc})^2}{(\sqrt{c}+\sqrt{d})^2}.\label{2-18-5}
\end{align}
holds, and this maximum value is attained when
$t_1=\frac{t \sqrt{d}}{\sqrt{c}+\sqrt{d}}$, 
$t_2=\frac{t \sqrt{c}}{\sqrt{c}+\sqrt{d}}$.
\end{lem}
\begin{proof}
Letting $x:=c t_1 +d t_2$, we have
$t_1= \frac{dt-x}{d-c}$ and 

$t_2= \frac{x-ct}{d-c}$.
Then, 
\begin{align*}
& a t_1+ b t_2
-\frac{( \sqrt{ac}t_1-\sqrt{bc}t_2)^2}{c t_1 +d t_2 }\\
=&
\left(\frac{\sqrt{ad}+\sqrt{bc}}{d-c}\right)^2
\left(
(c+d)t-x-\frac{cdt^2}{x}\right).
\end{align*}
Hence, the maximum is attained at $x= \sqrt{cd}t$,
{\it i.e.,}
$t_1=\frac{t \sqrt{d}}{\sqrt{c}+\sqrt{d}}$ and 
$t_2=\frac{t \sqrt{c}}{\sqrt{c}+\sqrt{d}}$.
Thus,
\begin{align*}
&\max_{t_1,t_2 \ge 0,~t_1+t_2= t}
a t_1+ b t_2
-\frac{( \sqrt{ac}t_1-\sqrt{bd}t_2)^2}{c t_1 +d t_2 }\\
=&
\left(\frac{\sqrt{ad}+\sqrt{bc}}{d-c}\right)^2
\left((c+d)t-2\sqrt{cd}t\right)
=
\frac{t(\sqrt{ad}+\sqrt{bc})^2}{(\sqrt{c}+\sqrt{d})^2}.
\end{align*}
\end{proof}
Further, three-parameter case can be maximized as follows.
\begin{lem}
The maximum value 
\begin{align*}
\max_{t_1,t_2,t_3 \ge 0,~t_1+t_2+t_3= t}
a t_1+ b t_2-\frac{( \sqrt{ac}t_1-\sqrt{bd}t_2)^2}{c t_1 +d t_2 +t_3}
\end{align*}
is equal to the maximum among three values
$\displaystyle\max_{t_1,t_3 \ge 0,~c t_1+t_3= t}a t_2-\frac{a c t_1^2}{c t_1 +t_3}$,
$\displaystyle\max_{t_2,t_3 \ge 0,~c t_2+t_3= t}a t_2-\frac{b d t_2^2}{d t_2 +t_3}$,
$\displaystyle\max_{t_1,t_2 \ge 0,~t_1+t_2= t}
a t_1+ b t_2-\frac{( \sqrt{ac}t_1-\sqrt{bd}t_2)^2}{c t_1 +d t_2 }$.
\end{lem}
\begin{proof}
Define two parameters 
$x:= c t_1 + d t_2 +t_3$ and 
$y:= \sqrt{cd}t_1-\sqrt{bd}t_2$.
Then, the range of $x$ and $y$ forms a convex set.
Since 
\begin{align*}
t_1 &= \frac{\sqrt{bd}(x-t) +(d-1)y }{\sqrt{bd}(c-1) +\sqrt{ac}(d-1)}, \\
t_2 &= \frac{\sqrt{ac}(x-t) -(c-1)y }{\sqrt{ac}(c-1) +\sqrt{bd}(d-1)}.
\end{align*}
Hence, 
\begin{align*}
&a t_1+ b t_2-\frac{( \sqrt{ac}t_1-\sqrt{bd}t_2)^2}{c t_1 +d t_2 +t_3}\\
=&
\biggl(
 \frac{a \sqrt{bd}}{\sqrt{bd}(c-1) +\sqrt{ac}(d-1)}\\
&\quad +\frac{b \sqrt{ac}}{\sqrt{ac}(c-1) +\sqrt{bd}(d-1)}
\biggr)(x-t)\\
&+\biggl(
 \frac{a(d-1) }{\sqrt{bd}(c-1) +\sqrt{ac}(d-1)} \\
&\quad -\frac{b(c-1) }{\sqrt{ac}(c-1) +\sqrt{bd}(d-1)}
\biggr)y
-\frac{y^2}{x}\\
=& -\frac{1}{x}(y-\frac{1}{2}Bx)^2 + (\frac{B^2}{4}+A)x-At,
\end{align*}
where
$A:=
 \frac{a \sqrt{bd}}{\sqrt{bd}(c-1) +\sqrt{ac}(d-1)}
+\frac{b \sqrt{ac}}{\sqrt{ac}(c-1) +\sqrt{bd}(d-1)}$,
$B:=
 \frac{a(d-1) }{\sqrt{bd}(c-1) +\sqrt{ac}(d-1)}
-\frac{b(c-1) }{\sqrt{ac}(c-1) +\sqrt{bd}(d-1)}$.
Applying Lemma \ref{2-18-6-a}, we obtain this lemma.
\end{proof}
\begin{lem}\label{2-18-6-a}
Define the function 
$f(x,y):= -\frac{1}{x}(y- \alpha x)^2+ \beta x$
on a closed convex set $C$.
The maximum value is realized at the boundary $bd C$.
\end{lem}
\begin{proof}
The condition can be classified to two cases:
i) $bd C \cap \{y=\alpha x\} = \emptyset $,
ii) $bd C \cap \{y=\alpha x\} \neq \emptyset $.
In the case i),
when fix $x$ is fixed, 
$\max_{y:(x,y)\in C}f(x,y)=
\max_{y:(x,y)\in bd C}f(x,y)$.
Then, we obtain 
$\max_{(x,y)\in C}f(x,y)=\max_{(x,y)\in bd C}f(x,y)$.
In the case ii), 
when $(x,\alpha x) \in C$,
$\max_{y:(x,y)\in C}f(x,y)=f(x,\alpha x)= \beta x$.
Hence,
$\max_{x:(x,\alpha x) \in C}
\max_{y:(x,y)\in C}f(x,y)= 
\max_{x:(x,\alpha x) \in C}\beta x$
This maximum is attained at
$x= \max_{x}\{x|:(x,\alpha x) \in C\}$ or 
$x= \min_{x}\{x|:(x,\alpha x) \in C\}$.
These point belongs to the boundary $bd C$.
Further,
$\max_{x:(x,\alpha x) \notin C}
\max_{y:(x,y)\in C}f(x,y)= 
\max_{x:(x,\alpha x) \in C}
\max_{y:(x,y)\in bd C}f(x,y)$.
Thus, the proof is completed.
\end{proof}

\section{Proof of Inequalities (\ref{2-24-7}) 
and (\ref{2-24-1-a})}\label{2-24-10}
It is sufficient to show 
\begin{align}
&\sqrt{\frac{3}{2}}
\left(
(2\fid+1)\sqrt{\lambda}+
\sqrt{6(\lambda (2\fid +1)+6 \delta)}
\right)\nonumber \\
& -
\Bigl(
(2\fid+1)
\sqrt{\lambda(2-2\fid)+6\delta} \nonumber \\
&\quad +
(2-2\fid)
\sqrt{\lambda(2\fid+1)+6\delta}
\Bigr)> 0.\label{2-24-22}
\end{align}
By putting $\delta':= \frac{6\delta}{\lambda}$,
the LHS is evaluated as
\begin{align*}
&\frac{\hbox{\rm LHS of }(\ref{2-24-22})}{\sqrt{\lambda}}\\
=&
\sqrt{\frac{3}{2}}
(2\fid+1)
+
3 \sqrt{(2\fid +1)+\delta')} \\
&-
(2\fid+1)
\sqrt{(2-2\fid)+\delta'}
-
(2-2\fid)
\sqrt{(2\fid+1)+\delta'} \\
=&
\sqrt{\frac{3}{2}}
(2\fid+1)
+
(2\fid+1)\sqrt{(2\fid +1)+\delta')}\\
& -
(2\fid+1)
\sqrt{(2-2\fid)+\delta'}\\
=&
(2\fid+1)
\left(
\sqrt{\frac{3}{2}}
+
\sqrt{(2\fid +1)+\delta')}
-
\sqrt{(2-2\fid)+\delta'}\right).
\end{align*}
Since $0\le \fid \le 1$,
we have
\begin{align*}
& \sqrt{\frac{3}{2}}
+
\sqrt{(2\fid +1)+\delta')}
-
\sqrt{(2-2\fid)+\delta'}\\
\ge &
\sqrt{\frac{3}{2}}
+
\sqrt{1+\delta'}
-
\sqrt{2+\delta'}.
\end{align*}
Further, the function $\delta' \to \sqrt{1+\delta'}
-\sqrt{2+\delta'}$ $(\delta' \in [0 , \infty])$ has 
the minimum $\sqrt{1}-\sqrt{2} > -1 > -\sqrt{\frac{3}{2}}$
at $\delta'=0$.
Hence, 
$\frac{\hbox{\rm LHS of }(\ref{2-24-22})}{\sqrt{\lambda}} > 0$.

\section{Proof of Equations (\ref{2-24-11}) and (\ref{2-24-3})}\label{2-24-12}
It is sufficient to show that
\begin{align}
&\sqrt{\frac{3}{2}}
\left(
(2-2\fid)\sqrt{\lambda}+
\sqrt{6(\lambda (2-2\fid )+6 \delta)}
\right)\nonumber \\
& -
\Bigl(
(2\fid+1)
\sqrt{\lambda(2-2\fid)+6\delta}\nonumber \\
&\quad +
(2-2\fid)
\sqrt{\lambda(2\fid+1)+6\delta}
\Bigr)>0 \label{2-24-4}
\end{align}
if and only if $\frac{6 \delta}{\lambda}< \delta_1$
and $\fid \ge \fid_{\frac{6 \delta}{\lambda}}$.
By putting $\delta':= \frac{6\delta}{\lambda}$,
the LHS of (\ref{2-24-4}) 
is evaluated as
\begin{align*}
&\frac{\hbox{\rm LHS of }(\ref{2-24-4})}{\sqrt{\lambda}}\\
=&
\sqrt{\frac{3}{2}}
(2-2\fid)
+
3 \sqrt{(2- 2\fid )+\delta')} \\
&-
(2\fid+1)
\sqrt{(2-2\fid)+\delta'}
-
(2-2\fid)
\sqrt{(2\fid+1)+\delta'} \\
=&
\sqrt{\frac{3}{2}}
(2-2\fid)
+
(2- 2\fid)
\sqrt{(2-2\fid)+\delta'}\\
& -
(2-2\fid)
\sqrt{(2\fid+1)+\delta'} \\
=&
(2-2\fid)
\left(
\sqrt{\frac{3}{2}}
+
\sqrt{(2-2\fid)+\delta'}
-
\sqrt{(2\fid+1)+\delta'} 
\right).
\end{align*}
Since $0\le \fid \le 1$
and $\delta \ge 0$,
\begin{align*}
\sqrt{\frac{3}{2}}
+
\sqrt{(2-2\fid)+\delta'}
-
\sqrt{(2\fid+1)+\delta'} >0
\end{align*}
if and only if
$\delta_1> \delta'$ and $\fid > \fid_{\delta'}$.

\section{Proof of (\ref{5-2-1})}\label{3-6-10}
Define $\tilde{\mu}_i$ by
\begin{align*}
\min_{\vec{w}\cdot \vec{\mu}'= c_0}
D({\rm Poi}(0,\ldots,0, \tilde{\mu}_i,0,\ldots, 0)
\|{\rm Poi}(\vec{\mu}'))=R.
\end{align*}
In fact, when $w_i a < c_0$,
\begin{align*}
&\min_{\mu_i'\ge 0:~\vec{w}\cdot \vec{\mu}'= c_0}
D({\rm Poi}(0,\ldots,0, a ,0,\ldots, 0)
\|{\rm Poi}(\vec{\mu}'))\\
=&\min_{\mu_i'\ge 0:~\vec{w}\cdot \vec{\mu}'= c_0}
\sum_{j=1}^m \mu_j' - a+ a \log \frac{a}{\mu_i'}\\
=&\min_{\alpha\ge 0,\beta\ge 0:~
\mu_i \alpha + \mu_M \beta= c_0}
\alpha + \beta
- a + a \log \frac{a}{\alpha}\\
=&
\left\{
\begin{array}{ll}
\frac{c_0}{w_i} - a+ a \log \frac{a w_i}{c_0}
& \hbox{ if } 
a \ge \frac{c_0(w_M-w_i)}{w_M w_i}
\\
\frac{c_0}{w_M} + a \log \frac{w_M-w_i}{w_M}
&\hbox{ if } 
a < \frac{c_0(w_M-w_i)}{w_M w_i} .
\end{array}
\right.
\end{align*}
This value is monotone decreasing concerning $a$.
When $a= \frac{c_0(w_M-w_i)}{w_M w_i}$,
this value is 
$\frac{c_0}{w_M} - 
\frac{c_0(w_M-w_i)}{w_i w_M}\log \frac{w_M-w_i}{w_M}$.
Hence, 
the value $\tilde{\mu}_i$
coincides with the the value $\tilde{\mu}_i$
defined by (\ref{2-6-12}) and (\ref{2-6-11}).

Thus, the relation (\ref{5-2-1}) follows from the relation
\begin{align*}
& \min_{\vec{w}\cdot \vec{\mu}'= c_0}
D({\rm Poi}(p_1 a_1,\ldots,p_m a_m)
\|{\rm Poi}(\vec{\mu}')) \\
\le &
\sum_{i=1}^m
p_i
\displaystyle \min_{\vec{w}\cdot \vec{\mu}'= c_0}
D({\rm Poi}(0,\ldots,0, a_i ,0,\ldots, 0)
\|{\rm Poi}(\vec{\mu}')).
\end{align*}
We choose $\vec{\mu}'_{i}$
such that 
$\min_{\vec{w}\cdot \vec{\mu}'= c_0}
D({\rm Poi}(0,\ldots,0, a_i ,0,\ldots, 0)
\|{\rm Poi}(\vec{\mu}'))
=
D({\rm Poi}(0,\ldots,0, a_i ,0,\ldots, 0)
\|{\rm Poi}(\vec{\mu}'_{i}))$.
Then, the above inequality follows from Lemma \ref{le-3-12} in the 
following way:
\begin{align*}
& \sum_{i=1}^m
p_i \min_{\vec{w}\cdot \vec{\mu}'= c_0}
D({\rm Poi}(0,\ldots,0, a_i ,0,\ldots, 0)
\|{\rm Poi}(\vec{\mu}')) \\
= &
\sum_{i=1}^m
p_i 
D({\rm Poi}(0,\ldots,0, a_i ,0,\ldots, 0)
\|{\rm Poi}(\vec{\mu}'_{i})) \\
\ge &
D({\rm Poi}(p_1 a_1,\ldots,p_m a_m)
\|{\rm Poi}(
\sum_{i=1}^m p_i \vec{\mu}'_{i})) \\
\ge &
\min_{\vec{w}\cdot \vec{\mu}'= c_0}
D({\rm Poi}(p_1 a_1,\ldots,p_m a_m)
\|{\rm Poi}(\vec{\mu}')) .
\end{align*}

\begin{lem}\label{le-3-12}
Any real number $0 \le p \le 1$
and any four sequence of positive numbers $(\mu_i)$,
$(\nu_i)$, $(\mu_i')$, and $(\nu_i')$
satisfy 
\begin{align*}
&p
(\sum_{i=1}^m (\mu_i- \nu_i)+ \sum_{i=1}^m \nu_i \log \frac{\nu_i}{\mu_i})\\
&+(1-p)
(\sum_{i=1}^m (\mu_i'- \nu_i')
+ \sum_{i=1}^m \nu_i' \log \frac{\nu_i'}{\mu_i'}) \\
\ge &
\sum_{i=1}^m (
(p\mu_i+(1-p)\mu_i')
- 
p\nu_i+(1-p)\nu_i')) \\
&+ 
\sum_{i=1}^m (p\nu_i +(1-p)\nu_i')
\log \frac{(p\nu_i +(1-p)\nu_i')}{(p\mu_i+(1-p)\mu_i')}.
\end{align*}
\end{lem}
\begin{proof}
It is sufficient to show
\begin{align*}
&p
(\sum_{i=1}^m \nu_i \log \frac{\nu_i}{\mu_i})
+(1-p)
(\sum_{i=1}^m \nu_i' \log \frac{\nu_i'}{\mu_i'}) \\
\ge &
\sum_{i=1}^m (p\nu_i +(1-p)\nu_i')
\log \frac{(p\nu_i +(1-p)\nu_i')}{(p\mu_i+(1-p)\mu_i')}.
\end{align*}
The convexity of $- \log $ implies that
\begin{align*}
&-\log(
\frac{(p\mu_i+(1-p)\mu_i')}{(p\nu_i +(1-p)\nu_i')}
)\\
=&
-\log(
\frac{p\nu_i}{(p\nu_i +(1-p)\nu_i')}\frac{\mu_i}{\nu_i}
+ \frac{(1-p)\nu_i'}{(p\nu_i +(1-p)\nu_i')}\frac{\mu_i'}{\nu_i'}
)\\
\le &
\frac{p\nu_i}{(p\nu_i +(1-p)\nu_i')}
\cdot -\log(
\frac{\mu_i}{\nu_i})\\
&+ 
\frac{(1-p)\nu_i'}{(p\nu_i +(1-p)\nu_i')}
\cdot -\log(\frac{\mu_i'}{\nu_i'}).
\end{align*}
Hence,
\begin{align*}
&p
(\sum_{i=1}^m \nu_i \log \frac{\nu_i}{\mu_i})
+(1-p)
(\sum_{i=1}^m \nu_i' \log \frac{\nu_i'}{\mu_i'}) \\
=&
\sum_{i=1}^m
(p\nu_i +(1-p)\nu_i')
\Bigl(\frac{p\nu_i}{(p\nu_i +(1-p)\nu_i')}
\cdot -\log(
\frac{\mu_i}{\nu_i})\\
&+ 
\frac{(1-p)\nu_i'}{(p\nu_i +(1-p)\nu_i')}
\cdot -\log(\frac{\mu_i'}{\nu_i'})\Bigr)\\
\ge & -\sum_{i=1}^m (p\nu_i +(1-p)\nu_i')
\log \frac{(p\nu_i +(1-p)\nu_i')}{(p\mu_i+(1-p)\mu_i')}.
\end{align*}
\end{proof}

\section{Proof of (\ref{5-5-1})}\label{a5}
Considering the shape of the graph $\frac{x}{\sqrt{y}}=a$,
we can show that
the minimum value $\min_{(x,y)\in C}\frac{x}{\sqrt{y}}$
can be attained by the boundary of $C$.
Hence the boundary of the convex set $Co(R)$ 
is included by the union 
$\cup_{i \neq j}l_{i,j}$
of the lines $l_{i,j}=\{
(t x_i(R)+(1-t)x_j(R),t y_i(R)+(1-t)y_j(R))|
0 \le t \le 1\}$.
Taking the derivative of 
$\frac{t x_i(R)+(1-t)x_j(R)}
{\sqrt{t y_i(R)+(1-t)y_j(R)}}$ concerning $t$,
we obtain 
\begin{align}
\min_{t \in [0,1]}
\frac{t x_i(R)+(1-t)x_j(R)}
{\sqrt{t y_i(R)+(1-t)y_j(R)}}
= z_{i,j}(R).
\end{align}
Hence, we obtain (\ref{5-5-1}).

\end{document}